\documentclass[11pt]{article}
\usepackage{cite}
\usepackage{amsmath,amsfonts,amssymb,graphicx,calrsfs}
\usepackage[small,bf,hang]{caption}
\usepackage{slashed}
\usepackage{mathabx,amsmath}
\usepackage{color}

\usepackage[vcentermath]{youngtab}
\usepackage{mathrsfs}

\usepackage{dsfont}
\usepackage[Symbol]{upgreek}

\def\hybrid{
        \topmargin -20pt
        \oddsidemargin 0pt
        \headheight 0pt \headsep 0pt
        \textwidth 6.25in 
        \textheight 9.5in 
        \marginparwidth .875in
        \parskip 5pt plus 1pt \jot = 1.5ex}

\hybrid

\usepackage{color}
\definecolor{red}{rgb}{1,0,0}
\definecolor{lred}{rgb}{0.3,0,0}
\definecolor{green}{rgb}{0,0.6,0}
\definecolor{blue}{rgb}{0,0,1}
\definecolor{violet}{rgb}{0.8,0,0.8}

\usepackage[usenames,dvipsnames]{xcolor}

\usepackage[colorlinks=true,
pdfstartview=FitV,
linkcolor= BrickRed,
citecolor= BrickRed,
urlcolor= BrickRed,
hyperindex=true,
hyperfigures=true]
{hyperref}
\hypersetup{linktoc=page}

 \csname
@addtoreset\endcsname{equation}{section}

\def\moth{\mathsurround=0pt}
\newdimen\zo \zo=0pt

\def\tick{\leaders\hrule height 0.5ex depth 0pt \hskip 0.5pt}
\def\upboxfill{$\moth \setbox\zo\hbox{\tick}%
  \hskip 3pt\hbox to 0pt{$\tick$\hss}\hrulefill \hbox to 7.5pt{$\tick$\hss}$}

\def\dtick{\leaders\hrule height .34pt depth 0.5ex \hskip 0.5pt}
\def\downboxfill{$\moth \setbox\zo\hbox{\dtick}%
  \hskip 2pt\hbox to 0pt{$\dtick$\hss}\hrulefill \hbox to 2pt{$\dtick$\hss}$}

\def\be#1\ee{\begin{align}#1\end{align}}

\def\bec{\begin{center}}
\def\ec{\end{center}}

\def\bea{\begin{eqnarray}}
\def\eea{\end{eqnarray}}
\def\ba{\begin{array}}
\def\ea{\end{array}}

\newcommand{\partialsix}{\partial_y }

\begin{document}

\begin{titlepage}
\rightline{}
\rightline{July 2020}
\rightline{HU-EP-20/08}
\begin{center}
\vskip 1.5cm
 {\LARGE \bf{ Toward Exotic 6D Supergravities}}
\vskip 1.7cm

{\large\bf {Yannick Bertrand$^a$, Stefan Hohenegger$^a$, Olaf Hohm$^b$, Henning Samtleben$^c$}}
\vskip .8cm

{\it  $^a$ Univ Lyon, Univ Claude Bernard Lyon 1, CNRS/IN2P3, \\
IP2I Lyon, UMR 5822, F-69622, Villeurbanne, France}\\
y.bertrand@ipnl.in2p3.fr, s.hohenegger@ipnl.in2p3.fr
\vskip .3cm

{\it  $^b$ Institute for Physics, Humboldt University Berlin,\\
 Zum Gro\ss en Windkanal 6, D-12489 Berlin, Germany}\\
 ohohm@physik.hu-berlin.de
\vskip .3cm

{\it  $^c$ Univ Lyon, Ens de Lyon, Univ Claude Bernard, CNRS,\\
Laboratoire de Physique, F-69342 Lyon, France} \\
{henning.samtleben@ens-lyon.fr}
\vskip .3cm

\vskip .2cm

\end{center}

\bigskip\bigskip
\begin{center} 
\textbf{Abstract}

\end{center} 
\begin{quote}

We investigate exotic supergravity theories in 6D with maximal $(4,0)$ and $(3,1)$ supersymmetry, 
which were conjectured by C.\ Hull to exist and to describe strong coupling limits of ${\cal N}=8$ theories in 5D. 
These theories involve exotic gauge fields with non-standard Young tableaux representations, 
subject to (self-)duality constraints. We give novel actions in a $5+1$ split of coordinates whose 
field equations reproduce those of the free bosonic $(4,0)$ and $(3,1)$ theory, respectively, including the 
(self-)duality relations. Evidence is presented for a master exceptional field theory formulation 
with an extended section constraint that, depending on the solution, produces the 
$(4,0)$, $(3,1)$ or the conventional $(2,2)$ theory. We comment on the possible construction of a fully 
non-linear master exceptional field theory.

\end{quote} 
\vfill
\setcounter{footnote}{0}
\end{titlepage}

\tableofcontents


\newpage

\section{Introduction}

Among the surprising  features of string/M-theory is the possible existence of exotic superconformal field and gravity theories 
 in six dimensions, which  display generalizations of electric-magnetic 
duality. Specifically, the supermultiplets of these theories are such that the corresponding fields must be subject to 
(self-)duality constraints, some of which involve exotic Young tableaux representations. 
In this paper our focus will be on a conjecture by Hull \cite{Hull:2000zn,Hull:2000rr} according to which there are  strong coupling limits of 
${\cal N}=8$ theories in five dimensions that are given by six-dimensional theories with chiral ${\cal N}=(4,0)$ and
${\cal N}=(3,1)$ supersymmetry, respectively. 
Such theories must be exotic or non-geometric since they feature mixed symmetry tensors of 
Young tableaux type $\,{\tiny \yng(2,1)}\;$ and ${\tiny \yng(2,2)}\;$,
respectively, instead of a conventional graviton, hence suggesting the need for a 
generalized notion of spacetime and diffeomorphism invariance. 
They are set to play a distinguished role among the maximally supersymmetric theories~\cite{Chiodaroli:2011pp,Anastasiou:2013hba,Borsten:2017jpt}

A possible window into these somewhat mysterious structures is offered by a Kaluza-Klein perspective from five dimensions.
The supermultiplets of five-dimensional (5D) theories with maximal supersymmetry (32 real supercharges) were classified by Strathdee \cite{Strathdee:1986jr} 
and further clarified by Hull in \cite{Hull:2000cf}. The 5D superalgebra reads 
 \bea\label{superAlgebra}
  \{Q_{\alpha}^{a}, Q_{\beta}^b\} = \Omega^{ab}(\Gamma^{\mu}C)_{\alpha\beta} P_{\mu}+C_{\alpha\beta}(Z^{ab}+\Omega^{ab}K)\,,
 \eea
where $\alpha,\beta,\ldots =  1,\ldots, 4$ are the space-time spinor indices 
and $a,b,\ldots=1,\ldots, 8$ are ${\rm USp}(8)$ $R$-symmetry indices. 
This superalgebra features 27 central charges $Z^{ab}$, satisfying $Z^{ab}=-Z^{ba}$, $\Omega^{ab}Z_{ab}=0$, and a singlet central charge $K$. 
The BPS multiplets of this superalgebra describe  the possible Kaluza-Klein towers that any six-dimensional (6D) theory with maximal 
supersymmetry displays when compactified on a circle. 
For the conventional maximal 6D supergravity, which features  ${\cal N}=(2,2)$ supersymmetry, 
the massive Kaluza-Klein states do not carry the singlet central charge $K$. Instead, they carry a particular central charge $Z^{ab}$
transforming as a singlet under the six-dimensional $R$-symmetry group ${\rm USp}(4)\times{\rm USp}(4)$\,.
In contrast, the massive multiplets of the exotic theories carry non-vanishing singlet charge $K$
(together with nonvanishing $Z^{ab}$, singlet under ${\rm USp}(4)\times{\rm USp}(4)$ 
in the case of the ${\cal N}=(3,1)$ multiplets)~\cite{Hull:2000cf}.
This points to a unifying framework in the spirit of exceptional field theories~\cite{Berman:2010is,Coimbra:2011ky,Hohm:2013pua,Hohm:2013vpa}
which we will elaborate on in this paper.

Exceptional field theory (ExFT) provides in particular a formulation of 11-dimensional (11D) and type IIB supergravity 
in a form that is covariant under the global symmetry group E$_{6(6)}$ of 5D maximal supergravity, thanks to  
extended coordinates in the ${\bf 27}$ representation of this group, which are added to the five coordinates of 5D supergravity. 
The resulting theory is thus based on a 
$(5+27)$-dimensional spacetime split, in which the 27 coordinates are subject to an E$_{6(6)}$ covariant `section constraint' restricting them 
to a  suitable physical subspace, from which the complete (untruncated) 11D supergravity can be reconstructed, albeit in a Kaluza-Klein type formulation 
with a $5+6$ split of coordinates. 
(Equivalently, one may think of this as coupling the infinite towers of massive Kaluza-Klein multiplets to 5D supergravity, 
which reconstructs the complete 11D supergravity.) 
A different physical section reproduces the IIB theory~\cite{Hohm:2013pua,Baguet:2015xha}.
From a higher-dimensional perspective, the extra coordinates can be thought of as accounting for the 
possible brane windings, which in turn are related to the 27 central charges of the supersymmetry algebra (\ref{superAlgebra}).
The structure of the exotic supermultiplets then suggests an inclusion of their couplings within a suitable extension of this
framework.

In this paper we will present actions for (the bosonic sectors of) the free exotic 6D theories that generalize 
the (linearized) E$_{6(6)}$ exceptional field theory  \cite{Hohm:2013pua,Hohm:2013vpa} 
by adding one more `exotic' coordinate to the ${27}$, as suggested by 
the singlet central charge in (\ref{superAlgebra}). 
As one of the most enticing outcomes of our investigation we find  evidence for a master exceptional field theory formulation 
in which the conventional ${\cal N}=(2,2)$ theory as well as the ${\cal N}=(4,0)$ and ${\cal N}=(3,1)$ theory are obtained 
through different solutions of 
an extended section constraint of the form 
 \bea\label{sectionIntro}
  d^{MNK}\partial_N\otimes \partial_K -\frac1{\sqrt{10}}\,\Delta^{MN}(\partial_N\otimes \partial_{\bullet}+\partial_{\bullet}\otimes \partial_{N}) = 0 \,,
  \label{section_intro}
 \eea
where $M,N=1,\ldots, 27,$ are fundamental  $E_{6(6)}$ indices, $d^{MNK}$ denotes the $E_{6(6)}$ invariant fully symmetric tensor, 
and $\partial_{\bullet}$ is the derivative dual to the exotic coordinate. Moreover, $\Delta^{MN}$ denotes 
the (constant) background part of the generalized metric ${\cal M}^{MN}$ encoding all scalar fields. 
The first term in equation (\ref{section_intro}) defines the section constraint of $E_{6(6)}$ exceptional field theory, 
whose solutions restrict to the standard $D=11$ and IIB sections.
The second term encodes the extension of the constraint allowing for two more solutions corresponding to ${\cal N}=(4,0)$ and ${\cal N}=(3,1)$, respectively. 
More precisely, we recover the ${\cal N}=(4,0)$ exotic theory by dropping all dependence on the 27 standard coordinates,
and keeping only the dependence on the exotic coordinate.
The ${\cal N}=(3,1)$ model in turn is recovered by superposing this coordinate with the ${\rm F}_{4(4)}$ singlet under $27\rightarrow 26+1$ 
among the 27 coordinates of ExFT.
While we give several independent pieces of evidence for the existence of this master formulation 
(some  of which entail highly non-trivial numerical agreement), we also point out some gaps of the master formulation as understood so far. 
This implies that the complete non-linear theory requires new ingredients, not the least of which is a section constraint 
that makes sense for the non-linear theory and 
that reduces to 
(\ref{sectionIntro})  in the appropriate limit. 

As a technical result, we present novel actions for the bosonic sectors of the ${\cal N}=(3,1)$ and the ${\cal N}=(4,0)$ model 
that are based on a 5+1 split of the six-dimensional space-time, sacrificing manifest 6D Poincar\'e invariance. In the spirit of ExFT, these are
two-derivative actions which upon dimensional reduction to five dimensions reduce to the same action of linearized
maximal 5D supergravity. All dual fields, in particular the entire dual graviton sector, only appear under derivative along 
the sixth dimension.
The full field equations obtained by variation combine the second order Fierz-Pauli equations with first-order duality equations
defining the dual graviton sector. 
Actions for selfdual fields based on a 5+1 split of spacetime date back to \cite{Henneaux:1988gg} with the 
description of selfdual 6D tensor fields. More recently, actions for the ${\cal N}=(3,1)$ and ${\cal N}=(4,0)$ models
have been constructed in~\cite{Henneaux:2016opm,Henneaux:2017xsb,Henneaux:2018rub}, 
based on the prepotential formalism developed in 
\cite{Henneaux:2004jw}  in the context of linearized gravity.  Introduction of prepotentials for the gauge fields 
adapted to their self-duality properties allows for the construction of an action of fourth order in spatial derivatives.
Our construction is closer in spirit to the original construction of \cite{Henneaux:1988gg}, 
albeit dual in a sense discussed in more detail in appendix~\ref{app:HTExFT}.
It provides a novel mechanism for describing self-dual exotic tensor fields.

The rest of this paper is organized as follows. In sec.~2 we review the bosonic sector of the
${\cal N}=(2,2)$, ${\cal N}=(3,1)$ and ${\cal N}=(4,0)$ theories at the level of the equations motion, 
which are manifestly 6D Lorentz invariant. In order to find  actions for 
these  theories  we abandon manifest Lorentz invariance by performing a 
$5+1$ split of coordinates in sec.~3 for each of these models. 
In sec.~4 we then present, as one of our main technical results, actions whose second-order 
Euler-Lagrange equations can be integrated in order to reproduce the correct dynamics of the three theories. 
In sec.~5 we present the master formulation as currently understood, highlight its successes, which strike us as significant, 
but also discuss the structural problems that remain. We close with a brief outlook. 

\textit{Note added:} While finalizing the present paper the preprint \cite{Minasian:2020vxn} appeared, which also 
investigates exotic theories in 6D.


\section{Review of 6D Models}

We study six-dimensional field theories in Minkowski space
with flat metric 
\bea
\eta_{\hat\mu\hat\nu}={\rm diag}\,\{-1,1,1,1,1,1\}
\;,
\qquad
\hat\mu, \hat\nu = 0, \dots, 5
\;.
\eea
The Poincar\'e algebra in six dimensions admits chiral $({\cal N}_+, {\cal N}_-)$ supersymmetric extensions
where ${\cal N}_\pm$ count the cumber of right- and left-handed supercharges, respectively~\cite{Strathdee:1986jr}.
In particular, maximal supersymmetry (32 real supercharges) allows for the three possibilities
$({\cal N}_+, {\cal N}_-)=(2,2)$, $({\cal N}_+, {\cal N}_-)=(3,1)$, and $({\cal N}_+, {\cal N}_-)=(4,0)$. 
The corresponding lowest-dimensional massless supermultiplets 
have the field content~\cite{Strathdee:1986jr}
\begin{subequations}
\bea
(2,2) &:&
(3,3;1,1)+(2,2;4,4)+(1,3;5,1)+(3,1;1,5)+(1,1;5,5) 
\nonumber\\
&&{}
+(2,3;4,1)+(3,2;1,4)+(2,1;4,5)+(1,2;5,4)
\;,\label{6Dmultiplets22}
\\[1ex]
(3,1) &:&
(4,2;1,1)+(2,2;14,1)+(3,1;6,2)+(1,1;14',2)
\nonumber\\
&&{}
+(4,1;1,2)+(3,2;6,1)+(2,1;14,2)+(1,2;14',1)
\;,\label{6Dmultiplets31}
\\[1ex]
(4,0) &:&
(5,1;1)+(3,1;27)+(1,1;42)+(4,1;8)+(2,1;48)\;,
\label{6Dmultiplets40}
\eea
\end{subequations}
organized into representations of the little group 
\bea
{\rm G}_0 &=&
{\rm SU}(2) \times {\rm SU}(2)  \times {\rm USp}(2{\cal N}_+)  \times {\rm USp}(2{\cal N}_-)
\;.
\eea
In this section, we briefly review the six-dimensional free theories associated to these multiplets.

\subsection{The ${\cal N}=(2,2)$ Model}

Let us start from the ${\cal N}=(2,2)$ multiplet corresponding to maximal supergravity in six dimensions.
Its bosonic field content comprises a metric, 25 scalar fields, 16 vectors, and 5 two-forms.
The full non-linear theory has been constructed in \cite{Tanii:1984zk} with the scalar fields parametrizing an
${\rm SO}(5,5)/\left({\rm SO}(5)\times {\rm SO}(5)\right)$ coset space.
For the purpose of this paper, we will only consider the linearized (free) theory with no couplings among the different
types of matter.

The linearized spin-2 sector carries the symmetric Pauli-Fierz field $h_{\hat\mu\hat\nu}$. With the linearized
Riemann tensor given by
\bea
R_{\hat\mu\hat\nu,\hat\rho\hat\sigma} &=& 
-\partial_{\hat\mu}\partial_{[\hat\rho} h_{\hat\sigma]\hat\nu}+\partial_{\hat\nu}\partial_{[\hat\rho} h_{\hat\sigma]\hat\mu}
\;,
\label{Riemann6D}
\eea
linearization of the Einstein-Hilbert Lagrangian gives rise to the massless Fierz-Pauli Lagrangian
\bea
{\cal L}_{h} &=&
-\frac12\,\partial_{\hat\mu} h^{\hat\mu\hat\nu}\,\partial_{\hat\nu} h_{\hat\rho}{}^{\hat\rho}
+\frac12\,\partial_{\hat\mu} h^{\hat\rho\hat\sigma} \, \partial_{\hat\rho} h_{\hat\sigma}{}^{\hat\mu}
-\frac14\,\partial_{\hat\mu} h^{\hat\rho\hat\sigma}\,\partial^{\hat\mu} h_{\hat\rho\hat\sigma}
+\frac14\,\partial_{\hat\mu} h_{\hat\nu}{}^{\hat\nu} \,\partial^{\hat\mu} h_{\hat\rho}{}^{\hat\rho}
\nonumber\\
&=& 
-\frac{1}{4}\,{\Omega}^{\hat\mu\hat\nu\hat\rho} {\Omega}_{\hat\mu\hat\nu\hat\rho}
+\frac12\,{\Omega}^{\hat\mu\hat\nu\hat\rho}{\Omega}_{\hat\nu\hat\rho\hat\mu}
+{\Omega}^{\hat\mu}{\Omega}_{\hat\mu}
\;,
\label{EH6}
\eea
with ${\Omega}_{\hat\mu\hat\nu\hat\rho}\equiv \partial_{[\hat\mu} h_{\hat\nu]\hat\rho}$\,.
The vector fields $A_{\hat\mu}{}^i$ couple with a standard Maxwell term
\bea
{\cal L}_A &=& -\frac14\,F_{\hat\mu\hat\nu}{}^i\,F^{\hat\mu\hat\nu\,i}
\;,
\qquad
i=1, \dots, 16\;,
\label{LA6}
\eea
for $F_{\hat\mu\hat\nu}{}^i=2\,\partial_{[\hat\mu} A_{\hat\nu]}{}^i$, while scalar couplings take the form
\bea
{\cal L}_\phi &=& -\frac12\,\partial_{\hat\mu} \phi^\alpha 
\partial^{\hat\mu} \phi^\alpha 
\;,
\qquad
\alpha=1, \dots, 25
\;.
\label{LP6}
\eea
The couplings (\ref{LA6}) and (\ref{LP6}) break the global ${\rm SO}(5,5)$ symmetry of
the non-linear theory down to its compact part ${\rm SO}(5)\times {\rm SO}(5)$, as expected for the free theory.
Finally, the two-forms $B_{\hat\mu\hat\nu}{}^p$ couple with a standard kinetic term
\bea
{\cal L}_B &=& -\frac16\,H_{\hat\mu\hat\nu\hat\rho}{}^q H^{\hat\mu\hat\nu\hat\rho\,q}
\;,\qquad
q=1, \dots, 5
\;,
\label{LB6}
\eea
for $H_{\hat\mu\hat\nu\hat\rho}{}^q=3\,\partial_{[\hat\mu} B_{\hat\nu\hat\rho]}{}^q$.
For the following it will be convenient to combine these fields together with their magnetic duals
into a set of 10 two-forms $B_{\hat\mu\hat\nu}{}^a$, satisfying first order (anti-)selfduality field equations
\bea
\delta_{ab}\,H_{\hat\mu\hat\nu\hat\rho}{}^b &=& 
\frac16\,\varepsilon_{\hat\mu\hat\nu\hat\rho\hat\sigma\hat\kappa\hat\lambda}\,\eta_{ab}\,H^{\hat\sigma\hat\kappa\hat\lambda\,b}
\;,
\qquad
a=1, \dots, 10
\;,
\label{6Dsd22}
\eea
with the ${\rm SO}(5,5)$ invariant constant tensor $\eta_{ab}$\,.
Equations (\ref{6Dsd22}) amount to a description of these degrees of freedom in terms of 5 selfdual and 5 anti-selfdual two forms.

\subsection{The ${\cal N}=(3,1)$ Model}

Let us now turn to the free field equations associated with the ${\cal N}=(3,1)$ multiplet (\ref{6Dmultiplets31}). This multiplet does not carry
a standard graviton field, but an exotic three-index tensor field of mixed-symmetry type \cite{Curtright:1980yk}
\bea
{\tiny \yng(2,1)} &:&
C_{\hat\mu\hat\nu,\hat\rho}=-C_{\hat\nu\hat\mu,\hat\rho}\;,\quad
C_{[\hat\mu\hat\nu,\hat\rho]}= 0
\;.
\label{C21}
\eea 
Its field equation is given by a selfduality equation~\cite{Hull:2000zn}
\bea
S_{\hat{\mu}\hat{\nu}\hat{\rho},\hat{\sigma}\hat{\tau}} &=& 
\frac16\, \varepsilon_{\hat{\mu}\hat{\nu}\hat{\rho}\hat{\eta}\hat{\kappa}\hat{\lambda}}\, S^{\hat{\eta}\hat{\kappa}\hat{\lambda}}{}_{\hat{\sigma}\hat{\tau}}\;,
\label{6DsdC31}
\eea
in terms of its second order curvature
\bea
S_{\hat{\mu}\hat{\nu}\hat{\rho},\hat{\sigma}\hat{\tau}} &=& 3\, \partial_{\hat{\sigma}} \partial_{[\hat{\mu}}C_{\hat{\nu}\hat{\rho}],\hat{\tau}}-3\, \partial_{\hat{\tau}} \partial_{[\hat{\mu}}C_{\hat{\nu}\hat{\rho}],\hat{\sigma}} \;.
\eea
Counting reveals that the field equation (\ref{6DsdC31}) captures the 8 degrees of freedom as counted in the multiplet (\ref{6Dmultiplets31}).
Moreover, curvature and field equation are invariant under the gauge symmetries
\bea
\delta C_{\hat{\mu}\hat{\nu},\hat{\rho}} &=& 2\,\partial_{[\hat{\mu}}\alpha_{\hat{\nu}]\hat{\rho}} + \partial_{\hat{\rho}} \beta_{\hat{\mu}\hat{\nu}} - \partial_{[\hat{\rho}}\beta_{\hat{\mu}\hat{\nu}]}\;,
\label{gauge31}
\eea
with parameters 
$\alpha_{\hat{\mu}\hat{\nu}} = \alpha_{(\hat{\mu}\hat{\nu})}$ and $\beta_{\hat{\mu}\hat{\nu}} = \beta_{[\hat{\mu}\hat{\nu}]}$.
An action principle for the field equations (\ref{6DsdC31}) has been constructed in \cite{Henneaux:2018rub}
based on the prepotential formalism introduced in \cite{Henneaux:2004jw} in the context of linearized gravity.

In addition to the exotic tensor field, the bosonic field content of the ${\cal N}=(3,1)$ multiplet (\ref{6Dmultiplets31}) contains 14 vectors, 12 selfdual 2-forms
and 28 scalar fields.
The dynamics of vector and scalar fields can be captured by standard Lagrangians (\ref{LA6}) and (\ref{LP6})
(with different range of internal indices).
The selfdual 2-forms $B_{\hat\mu\hat\nu}{}^a$ obey a selfduality equation similar to (\ref{6Dsd22})
\bea
H_{\hat\mu\hat\nu\hat\rho}{}^a &=& 
\frac16\,\varepsilon_{\hat\mu\hat\nu\hat\rho\hat\sigma\hat\kappa\hat\lambda}\,H^{\hat\sigma\hat\kappa\hat\lambda\,a}
\;,
\qquad
a=1, \dots, 12
\;,
\label{6Dsd31}
\eea
contrary to (\ref{6Dsd22}), no indefinite tensor $\eta_{ab}$ appears in this equation, all forms are selfdual.\footnote{
For uniformity, we use the same indices $a, b,$ to label two-forms in all three models, despite the fact that the range of these indices
differs among the different models according to the number of two-form fields.
This should not be a source of confusion.}
As a consequence there is no standard action principle for these field equations, they can however be derived
from an action with non-manifest Lorentz invariance \cite{Henneaux:1988gg} or upon coupling to the auxiliary PST scalar~\cite{Pasti:1996vs}.\footnote{
For more recent constructions, see also \cite{Sen:2019qit}, \cite{Mkrtchyan:2019opf}.}

The free ${\cal N}=(3,1)$ theory is invariant under the $R$-symmetry group ${\rm USp}(6)  \times {\rm USp}(2)$.
The (yet elusive) interacting theory is conjectured to exhibit a global ${\rm F}_{4(4)}$ symmetry
with in particular the 28 scalars parametrizing the coset space
${\rm F}_{4(4)}/\left({\rm USp}(6)  \times {\rm USp}(2)\right)$ \cite{Hull:2000zn}.

\subsection{The ${\cal N}=(4,0)$ Model}

The ${\cal N}=(4,0)$ multiplet carries an exotic four-index tensor field with the symmetries of the Riemann tensor
\bea
{\tiny \yng(2,2)} &:&
T_{\hat\mu\hat\nu,\hat\rho\hat\sigma}=T_{\hat\rho\hat\sigma,\hat\mu\hat\nu}=-T_{\hat\nu\hat\mu,\hat\rho\hat\sigma}\;,\quad
T_{[\hat\mu\hat\nu,\hat\rho]\hat\sigma}= 0
\;.
\label{T40}
\eea 
Its field equation is given by a selfduality equation~\cite{Hull:2000zn}
\bea
G_{\hat{\mu}\hat{\nu}\hat{\lambda},\hat{\rho}\hat{\sigma}\hat{\tau}} &=& 
\frac16\,\varepsilon_{\hat{\mu}\hat{\nu}\hat{\lambda}\hat{\alpha}\hat{\beta}\hat{\gamma}}\,
G^{\hat{\alpha}\hat{\beta}\hat{\gamma}}{}_{\hat{\rho}\hat{\sigma}\hat{\tau}}
\;,
\label{6DsdT40}
\eea
in terms of its second order curvature
\bea
G_{\hat{\mu}\hat{\nu}\hat{\lambda},\hat{\rho}\hat{\sigma}\hat{\tau}} &=& 
3\, \partial_{\hat{\rho}} \partial_{[\hat{\mu}}T_{\hat{\nu}\hat{\lambda}],\hat{\sigma}\hat{\tau}}
+3\, \partial_{\hat{\sigma}} \partial_{[\hat{\mu}}T_{\hat{\nu}\hat{\lambda}],\hat{\tau}\hat{\rho}}
+3\, \partial_{\hat{\tau}} \partial_{[\hat{\mu}}T_{\hat{\nu}\hat{\lambda}],\hat{\rho}\hat{\sigma}}
 \;.
\eea
Counting confirms that this field equation describes the 5 degrees of freedom as counted in the multiplet (\ref{6Dmultiplets40}).
Moreover, curvature and field equation are invariant under the gauge symmetries
\bea
\delta T_{\hat{\mu}\hat{\nu},\hat{\rho}\hat{\sigma}} &=& 
\partial_{[\hat{\mu}} \lambda_{\hat{\nu}],\hat{\rho}\hat{\sigma}} + 
\partial_{[\hat{\rho}} \lambda_{\hat{\sigma}],\hat{\mu}\hat{\nu}} 
\;,
\label{gauge40}
\eea
with the (2,1) gauge parameter $\lambda_{\hat{\mu},\hat{\rho}\hat{\sigma}}=\lambda_{\hat{\mu},[\hat{\rho}\hat{\sigma}]}$, 
$\lambda_{[\hat{\mu},\hat{\rho}\hat{\sigma}]}=0$\,.
An action principle for (\ref{6DsdT40}) has been constructed in \cite{Henneaux:2016opm,Henneaux:2017xsb}
based on the prepotential formalism of \cite{Henneaux:2004jw}.
The bosonic part of the ${\cal N}=(4,0)$ multiplet (\ref{6Dmultiplets40}) 
combines the exotic tensor field $T_{\hat\mu\hat\nu,\hat\rho\hat\sigma}$ with 42 scalars and 27 selfdual 2-forms.
Their dynamics is described by a free Lagrangian (\ref{LP6}) and selfduality equations (\ref{6Dsd31}), respectively.

The free ${\cal N}=(4,0)$ theory is invariant under the $R$-symmetry group ${\rm USp}(8)$.
The (yet elusive) interacting theory is conjectured to exhibit a global ${\rm E}_{6(6)}$ symmetry
with in particular the 42 scalars parametrizing the coset space
${\rm E}_{6(6)}{\rm USp}(8)$ \cite{Hull:2000zn}.

\section{$5+1$ Split}

Upon dimensional reduction to $D=5$ dimensions, the
three models discussed in the previous section all reduce to the same theory:
the free limit of maximal $D=5$ supergravity~\cite{Hull:2000zn,Hull:2000rr}.
The bosonic sector of this theory carries a spin-2 field and 27 vector fields together with 42 scalar fields.
In particular, the exotic tensor fields of the ${\cal N}=(3,1)$ and the ${\cal N}=(4,0)$ model after dimensional reduction
carry the $D=5$ dual graviton and double dual graviton, respectively.
Within the free theory, these fields can be dualized into the standard 
Pauli-Fierz field~\cite{Hull:2000zn,West:2001as,Hull:2001iu}, and do not represent independent degrees of freedom.
In order to make the equivalence explicit, the fields of $D=5$ supergravity (together with their on-shell duals)
have to be properly identified among the various components of the $D=6$ fields.

In this section, we discuss for every of the three models the reorganization of the $D=6$ fields which allows their identification 
after reduction to five dimensions. However, throughout this section (and this paper) we keep the full dependence of
all fields on six space-time coordinates.
More precisely, we break 6-dimensional Poincar\'e invariance  down to $5+1$ and perform a standard Kaluza-Klein decomposition
on the six-dimensional fields without dropping the dependence on the 6th coordinate. 
We then rearrange the equations such that they take the form of the five-dimensional (free) supergravity equations
however sourced by derivatives of matter fields along the sixth direction.
The resulting reformulation of the six-dimensional models casts their dynamics into a common framework --- which ultimately
allows us to construct uniform actions for the three models.

For the purpose of this paper, we choose the $5+1$ coordinate split 
\bea
\big\{ x^{\hat\mu} \big\} &\longrightarrow&
\left\{ x^{\mu}, y \right\}
\;,\qquad
\mu = 0, \dots, 4
\;,
\label{split51}
\eea
by singling out one of the spatial coordinates. 
Of course, an analogous construction
can be performed with a split along the time-like coordinate which may be of interest for example in a Hamiltonian context.

\subsection{The ${\cal N}=(2,2)$ Model}

With the coordinate split (\ref{split51}), we parametrize the graviton of the ${\cal N}=(2,2)$ theory as
\bea
h_{\hat\mu\hat\nu} &=& 
\begin{pmatrix}
h_{\mu\nu} -\frac13\,\eta_{\mu\nu}\,\phi & A_\mu \\
A_\mu & \phi
\end{pmatrix}
\;,
\eea
which is the linearized form of the standard Kaluza-Klein reduction ansatz. Recall that all fields still depend on 6 coordinates.
Working out the Lagrangian (\ref{EH6}) in this parametrization gives rise to its expression
\begin{align}
{\cal L}_{h} \;\longrightarrow\; {\cal L}_{\,{\Yboxdim{4pt}\yng(2)}} \;=\;&   -\frac12\,\partial_\mu h^{\mu\nu} \partial_\nu h_\rho{}^\rho  +\frac12\, \partial_\mu h^{\rho\sigma}\partial_\rho h_\sigma{}^\mu 
- \frac14\, \partial_\mu h^{\nu\rho}\partial^\mu h_{\nu\rho} + \frac14\, \partial_\mu h_\nu{}^\nu \partial^\mu h_\rho{}^\rho 
 \nonumber\\
&  - \frac14\, \partialsix h^{\mu\nu}\partialsix  h_{\mu\nu}+\partialsix  h^{\mu\nu} \partial_\mu A_\nu  -  \partialsix  h_\sigma{}^\sigma\partial^\rho A_\rho  
+ \frac{1}{4}\,\partialsix  h_\sigma{}^\sigma \partialsix  h_\rho{}^\rho  - \frac{2}{3}\, \partialsix  h_\sigma{}^\sigma  \partialsix  \phi 
 \nonumber \\
&- \frac{1}{4}\, F^{\mu\nu}F_{\mu\nu}
-  \frac{1}{3}\, \partial^\mu \phi \partial_\mu \phi  + \frac{4}{3}\, \partialsix  A^\mu  \partial_\mu \phi  + \frac{5}{9}\, \partialsix  \phi \partialsix  \phi
\;,
\label{L22h}
\end{align}
up to total derivatives. As an illustration of the above discussion 
let us note the explicit form of the equations for the five-dimensional spin-2 field
\bea
{\cal G}_{\mu\nu} 
 &=& 
-\frac12\, \partialsix  \partialsix  h_{\mu\nu}
+ \frac12\,\eta_{\mu\nu}\,\partialsix  \partialsix  h_\rho{}^\rho - \frac{2}{3}\,\eta_{\mu\nu}\, \partialsix  \partialsix  \phi
\;,
\label{Einstein22}
\eea
in terms of the linearized Einstein tensor
\bea
{\cal G}_{\mu\nu} &=& 
-\partial^\rho D_{(\mu} h_{\nu)\rho}
+\frac12\,\partial_\rho D^{\rho} h_{\mu\nu}
+\frac12\,\partial_{(\mu} D_{\nu)} h_\rho{}^\rho
+\frac12\,\eta_{\mu\nu}\,\partial^\rho D^\sigma h_{\rho\sigma}
-\frac12\,\eta_{\mu\nu}\,\partial_\rho D^{\rho} h_{\sigma}{}^\sigma
\;,
\nonumber\\
&&
\mbox{with covariant derivatives}\quad
D_{\mu} h_{\nu\rho} \equiv \partial_\mu  h_{\nu\rho}  - \frac23\,\partialsix  A_\mu\,\eta_{\nu\rho}
\;.
\label{GcovDh}
\eea
The form of (\ref{Einstein22}) shows that upon dimensional reduction to $D=5$ dimensions, these equations
reproduce the (linearized) five-dimensional Einstein field equations. In contrast, the coordinate dependence along
the sixth coordinate induces a non-trivial gauge structure via covariant derivatives (\ref{GcovDh}) and
non-vanishing source terms in (\ref{Einstein22}).
This is very much in the spirit of the reformulation of higher-dimensional supergravities as exceptional
field theories (ExFTs). Indeed, equation (\ref{Einstein22}) can be equivalently obtained upon linearizing the corresponding
E$_{6(6)}$ ExFT~\cite{Hohm:2013pua,Hohm:2013vpa} upon proper identification of the coordinate $y$ among the 27 internal
coordinates on which this ExFT is based.

Let us also note, that the Lagrangian (\ref{L22h}) can be put to the more compact form
\begin{align}
{\cal L}_{\,{\Yboxdim{4pt}\yng(2)}}=&   -\frac{1}{4}\,{\Omega}^{\mu\nu\rho} {\Omega}_{\mu\nu\rho}
+\frac12\,{\Omega}^{\mu\nu\rho}{\Omega}_{\nu\rho\mu}
+{\Omega}^{\mu}{\Omega}_{\mu}
-  \frac{1}{3}\, (\partial^\mu \phi -2\,\partialsix  A^\mu) (\partial_\mu \phi  -2\,\partialsix  A_\mu  ) 
\nonumber\\
&  
- \frac{1}{4}\, F^{\mu\nu}F_{\mu\nu} 
+ \frac{5}{9}\, \partialsix  \phi \partialsix  \phi
 - \frac{2}{3}\, \partialsix  h_\sigma{}^\sigma  \partialsix  \phi 
+ \frac{1}{4}\,\partialsix  h_\sigma{}^\sigma \partialsix  h_\rho{}^\rho 
- \frac14\, \partialsix h^{\mu\nu}\partialsix  h_{\mu\nu}
\;,
\label{L22h0}
\end{align}
with the linearized (and covariantized) anholonomity objects
\bea
{\Omega}_{\mu\nu\rho}&\equiv& \partial_{[\mu} h_{\nu]\rho} - \frac23\,\partialsix  A_{[\mu}{} \eta_{\nu]\rho} 
\;,
\qquad
\Omega_\mu ~\equiv~ \Omega_{\mu\nu}{}^\nu
\;.
\label{covOmega}
\eea

The remaining part of the six-dimensional degrees of freedom described by (\ref{L22h0})
are captured by a (modified) five-dimensional Maxwell and Klein-Gordon equation for $A_\mu$ and $\phi$, respectively,
obtained by varying (\ref{L22h0}).
It is useful to note the symmetries of the Lagrangian (\ref{L22h0}) descending from six-dimensional spin-2 gauge transformations
\bea
\delta h_{\mu\nu} &=& 2\,\partial_{(\mu} \xi_{\nu)} +\frac23\,\eta_{\mu\nu} \,\partialsix  \lambda
\;,\nonumber\\
\delta A_\mu &=& \partial_\mu \lambda + \partialsix  \xi_\mu 
\;,\nonumber\\
\delta \phi &=& 2\, \partialsix  \lambda
\;,
\eea
upon decomposition of the six-dimensional gauge parameter as $\{\xi^{\hat\mu}\} = \{\xi^\mu, \lambda\}$\,.

In a similar way, the six-dimensional Maxwell and Klein-Gordon Lagrangians
(\ref{LA6}) and (\ref{LP6}) take the form
\bea 
{\cal L}_A &=& -\frac{1}{4}\, F^{\mu\nu\,i}F_{\mu\nu}{}^i 
 - \frac{1}{2} \left(\partial^\mu \phi^i -\partialsix  A^{\mu\,i} \right)
 \left(\partial_\mu \phi^i -\partialsix  A_\mu{}^{i} \right)\;,
 \nonumber\\
{\cal L}_\phi &=&- \frac{1}{2}\,\partial^\mu \phi^\alpha \partial_\mu \phi^\alpha- \frac{1}{2}\,\partialsix  \phi^\alpha \partialsix  \phi^\alpha
\;,
\label{LAP22}
\eea
respectively, after splitting $\{A_{\hat\mu}{}^i\} = \{A_\mu{}^i, \phi^i\}$, and with abelian $F_{\mu\nu}{}^i=2\,\partial_{[\mu}A_{\nu]}{}^i$,
giving rise to modified Maxwell and Klein-Gordon equations for their components.
The rewriting of the tensor field sector is slightly less straightforward: rather than evaluating the Lagrangian (\ref{LB6}),
we choose to evaluate the first-order field equations (\ref{6Dsd22}) after splitting the 6D tensor fields into $\{B_{\hat\mu\hat\nu}{}^a\}
=\{B_{\mu\nu}{}^a, B_{\mu 5}{}^a \equiv A_\mu{}^a\}$
\bea
\eta_{ab}\,H_{\mu\nu\rho}{}^b &=& 
\frac12\,\varepsilon_{\mu\nu\rho\kappa\lambda}\,\delta_{ab}\,
(F^{\kappa\lambda\,b}+\partialsix  B^{\kappa\lambda\,b})
\;,
\label{sd22KK5}
\eea
where we use conventions $\varepsilon_{\mu\nu\rho\kappa\lambda5}=\varepsilon_{\mu\nu\rho\kappa\lambda}$,
and abelian field strengths $H_{\mu\nu\rho}{}^a=3\,\partial_{[\mu} B_{\nu\rho]}{}^a$, and 
$F_{\mu\nu}{}^a=2\,\partial_{[\mu} A_{\nu]}{}^a$, respectively.
These equations can be integrated to a Lagrangian
\bea
{\cal L}_{\,{\Yboxdim{4pt}\yng(1,1)}} &=& 
-\frac14\,\left( F_{\mu\nu}{}^a+ \partialsix  B_{\mu\nu}{}^a \right)\left( F^{\mu\nu\,a}+ \partialsix  B^{\mu\nu\,a} \right)-
\frac1{24}\,\varepsilon^{\mu\nu\rho\sigma\tau}\,\eta_{ab}\, \partialsix  B_{\mu\nu}{}^a \,
H_{\rho\sigma\tau}{}^b
\;.
\label{LBB22}
\eea
Again, this Lagrangian can be deduced from the linearized version of exceptional field theory. 
We discuss this mechanism in more detail in appendix~\ref{app:ExFT}.
As we will see in the following, 
this form of the Lagrangian allows for the most uniform treatment of the different six-dimensional models.
After dimensional reduction to $D=5$ dimensions, it simply reduces to a collection of Maxwell terms,
such that all degrees of freedom of (\ref{LB6}) are described as massless vector fields in five dimensions.
In presence of the sixth dimension, the Lagrangian (\ref{LBB22}) gives rise to modified Maxwell equations
while variation w.r.t.\ the tensor fields $B_{\mu\nu}{}^a$ induces equations (\ref{sd22KK5}) (under $\partialsix $ derivative) 
as duality equations relating vector and tensor fields.

In summary, the $D=6$ ${\cal N}=(2,2)$, model can be equivalently reformulated in terms of a Lagrangian given by the sum 
of (\ref{L22h0}), (\ref{LAP22}), and (\ref{LBB22}). Upon dimensional reduction to five dimensions,
i.e.\ setting $\partialsix  \rightarrow0$, and rescaling of the scalar fields, this Lagrangian reduces to
\begin{align}
{\cal L}_{5D}=&   
  -\frac{1}{4}\,{\mathring\Omega}^{\mu\nu\rho} {\mathring\Omega}_{\mu\nu\rho}
+\frac12\,{\mathring\Omega}^{\mu\nu\rho}{\mathring\Omega}_{\nu\rho\mu}
+{\mathring\Omega}^{\mu}{\mathring\Omega}_{\mu}
-\frac12\,\partial^\mu \phi^A \partial_\mu \phi^A
-\frac{1}{4}\, F^{\mu\nu\,M}F_{\mu\nu}{}^M 
- \frac14\, \partialsix  h^{\mu\nu}\partialsix  h_{\mu\nu} 
\;,
\nonumber\\
& 
\qquad\quad
M=1, \dots, 27\;,\quad
A=1, \dots, 42\;,
\label{5D22}
\end{align}
with $\mathring{\Omega}_{\mu\nu\rho}\equiv\partial_{[\mu}h_{\nu]\rho}$, and
where we have combined the various vector and scalar fields into joint objects 
\bea
\left\{A_\mu{}^M\right\} \;,\;\; M=1, \dots, 27\;,\qquad
\left\{\phi^A\right\} \;,\;\; A=1, \dots, 42
\;.
\label{AP5}
\eea
The Lagrangian (\ref{5D22}) is the free limit of $D=5$ maximal supergravity \cite{Cremmer:1980gs}.
In the interacting theory, the fields (\ref{AP5}) transform in the fundamental and a non-linear representation
of its global symmetry group E$_{6(6)}$.

\subsection{The ${\cal N}=(3,1)$ Model}
\label{subsec:split31}

We now turn to the ${\cal N}=(3,1)$ model. Its most characteristic element is the mixed-symmetry tensor field $C_{\hat\mu\hat\nu,\hat\rho}$
whose field equation (\ref{6DsdC31}) cannot be derived from a standard action principle. We thus perform the Kaluza-Klein
reorganization of the model on the level of the field equations.
To this end, we again split coordinates as (\ref{split51}) and parametrize the mixed-symmetry tensor as
\bea
\left\{C_{\hat\mu\hat\nu,\hat\rho}
\right\} &=&
\left\{ C_{\mu\nu,\rho}-2\,A_{[\mu}\eta_{\nu]\rho}\,;\; 
C_{\mu5,\nu} = h_{\mu\nu}+B_{\mu\nu}\,;\; 
C_{\mu5,5} = 2\,A_\mu \right\}
\;,
\label{CAB31}
\eea
with symmetric $h_{\mu\nu}=h_{\nu\mu}$, antisymmetric $B_{\mu\nu}=-B_{\nu\mu}$,
and a (2,1) tensor $C_{\mu\nu,\rho}$\,.
After dimensional reduction to five dimensions, the fields $h_{\mu\nu}$ and $A_\mu$ satisfy
the linearized Einstein and Maxwell equations while the fields $C_{\mu\nu,\rho}$ and $B_{\mu\nu}$
describe their on-shell duals, together accounting for the 8 degrees of freedom of the six-dimensional tensor field.
Explicitly, in the parametrization (\ref{CAB31}), the six-dimensional selfduality equations (\ref{6DsdC31}) split into two equations
\bea
\partial_\rho\Big(
F_{\mu\nu}+\frac16\,\varepsilon_{\mu\nu\lambda\sigma\tau}\,H^{\lambda\sigma\tau}
\Big)&=& 
\partialsix \partial_{[\mu}h_{\nu]\rho}
+\frac14\,\varepsilon_{\mu\nu\kappa\lambda\tau}\,\partialsix  \partial^\kappa C^{\lambda\tau}{}_\rho
+\frac12\,\partialsix  \partialsix  C_{\mu\nu,\rho} 
-\partialsix  \partialsix  A_{[\mu}\eta_{\nu]\rho}
\nonumber\\
&&{}
-\frac14\,\varepsilon_{\rho\mu\nu\sigma\tau}\,\partialsix  F^{\sigma\tau}
+\partialsix \partial_{[\mu}B_{\nu]\rho} -\partialsix  \partial_\rho B_{\mu\nu}
\;,
\label{eomhC}
\eea
and
\bea
R_{\mu\nu,\rho\sigma}
&=&
\frac12\,\partial_\rho \Big(
H_{\mu\nu\sigma}
-\frac12\,\varepsilon_{\mu\nu\sigma\kappa\lambda}\, F^{\kappa\lambda}  \Big)
-\frac12\,\partial_\sigma  \Big(
H_{\mu\nu\rho}
-\frac12\,\varepsilon_{\mu\nu\rho\kappa\lambda}\,  F^{\kappa\lambda}  \Big)
+\frac12\,\varepsilon_{\mu\nu\kappa\lambda\tau}\,\partial^\kappa \partial_{[\rho}C^{\lambda\tau}{}_{\sigma]}
 \nonumber\\
&&{}
+\frac12\,\partialsix  \partial_\rho C_{\mu\nu,\sigma}
-\frac12\,\partialsix  \partial_\sigma C_{\mu\nu,\rho}
-\partialsix  \partial_\rho A_{[\mu} \eta_{\nu]\sigma}
+\partialsix  \partial_\sigma A_{[\mu} \eta_{\nu]\rho}
\;,
\label{eomR6}
\eea
with abelian field strengths $F_{\mu\nu}=2\,\partial_{[\mu}A_{\nu]}$, $H_{\mu\nu\rho}=3\,\partial_{[\mu} B_{\nu\rho]}$, and
the linearized Riemann tensor $R_{\mu\nu,\rho\sigma}$ defined as in (\ref{Riemann6D}) however 
for the field $h_{\mu\nu}$\,.
Contraction of (\ref{eomR6}) gives rise to an equation
\bea
{\cal G}_{\mu\nu}
&=&
-\frac12\,\partialsix  \partial^\rho C_{\rho(\mu,\nu)}
-\frac12\,\partialsix \partial_{(\mu} C_{\nu)\rho}{}^\rho
+\frac12\,\eta_{\mu\nu}\,\partialsix  \partial^\rho C_{\rho\sigma}{}^\sigma
\;,
\label{Einstein31}
\eea
where ${\cal G}_{\mu\nu}$ denotes the linearized Einstein tensor defined as in (\ref{GcovDh}),
however with covariant derivatives now given by
\bea
D_{\mu} h_{\nu\rho} \equiv \partial_\mu  h_{\nu\rho}  - \partialsix  A_\mu\,\eta_{\nu\rho}
\;,
\label{covDh31}
\eea
i.e.\ with a different value of the coupling constant (which could be absorbed into rescaling the vector field).
Equation (\ref{Einstein31}) confirms that upon reduction to five dimensions ($\partialsix \rightarrow0$), the field $h_{\mu\nu}$
satisfies the linearized Einstein equations. As in the ${\cal N}=(2,2)$ model, the coordinate dependence along the sixth coordinate
induces a nontrivial gauge structure (\ref{covDh31}) together with non-vanishing source terms in (\ref{Einstein31}) ---
which differ from those of (\ref{Einstein22}) illustrating the inequivalence of the ${\cal N}=(2,2)$ and the ${\cal N}=(3,1)$ model before dimensional reduction.

The full field equation (\ref{eomR6}) takes the form of a vanishing curl (in $[\rho\sigma]$) and can locally 
be integrated into the first order equation\footnote{
Here, and in the following we work locally and 
ignore potential subtleties that may arise from a non-trivial topology. 
We refer to \cite{Bekaert:1998yp} for a discussion of such issues in the context of chiral $p$-forms.
}
\bea
\partial_{[\mu} h_{\nu]\rho}
+\frac1{4}\,\varepsilon_{\mu\nu\kappa\lambda\tau}\,\partial^\kappa C^{\lambda\tau}{}_{\rho}
+\frac12\,
\Big(
H_{\mu\nu\rho}
-\frac12\,\varepsilon_{\mu\nu\rho\kappa\lambda}\,  F^{\kappa\lambda} \Big)
+\frac1{2}\,\partialsix   C_{\mu\nu,\rho}
-\partialsix   A_{[\mu} \eta_{\nu]\rho} 
&=& \partial_\rho u_{\mu\nu}
\;,
\nonumber\\
\label{defc}
\eea
with an antisymmetric tensor $u_{\mu\nu}=-u_{\nu\mu}$\,.
Combining this equation with the field equation (\ref{eomhC}) implies that
\bea
\partial_\rho\left(
F_{\mu\nu}
+\frac16\,\varepsilon_{\mu\nu\kappa\lambda\tau}\,H^{\kappa\lambda\tau}
+\frac32\,\partialsix  B_{\mu\nu}  -\partialsix  u_{\mu\nu} 
\right)
&=&0
\;,
\label{dualFH6der}
\eea
which can be further integrated into another first order duality equation
\bea
F_{\mu\nu}
+\frac16\,\varepsilon_{\mu\nu\kappa\lambda\tau}\,H^{\kappa\lambda\tau}
+\frac32\,\partialsix  B_{\mu\nu}  -\partialsix  u_{\mu\nu} 
&=& 0
\;,
\label{dualFH6}
\eea
up to a function $f_{\mu\nu}(y)$ that can be absorbed into $u_{\mu\nu}$\,.
Eventually, we can use (\ref{dualFH6}) to bring (\ref{defc}) into the form
\bea
\partial_{[\mu} h_{\nu]\rho}
+\frac1{4}\,\varepsilon_{\mu\nu\kappa\lambda\tau}\,\partial^\kappa C^{\lambda\tau}{}_{\rho}
 -\partial_\rho u_{\mu\nu}
 &=&
\frac14\,\varepsilon_{\mu\nu\rho\kappa\lambda}\,  
\partialsix  \Big(
u^{\kappa\lambda} -
\frac32\, B^{\kappa\lambda}  
\Big)
-\frac1{2}\,\partialsix   C_{\mu\nu,\rho}
+\partialsix   A_{[\mu} \eta_{\nu]\rho} 
\;.
\nonumber\\
\label{defc2}
\eea
To sum up, we have cast the original second order field equations (\ref{6DsdC31}) 
of the six-dimensional mixed-symmetry tensor field into the form of two first-order
duality equations (\ref{dualFH6}) and (\ref{defc2}), upon parametrizing the six-dimensional
fields in terms of its components (\ref{CAB31}) and introduction of an additional field $u_{\mu\nu}$\,.
Upon reduction to five dimensions, these equations constitute the duality equations
relating the vector-tensor fields, and the graviton-dual graviton fields, respectively.

It is instructive to work out the gauge symmetries of these equations which originate 
from the $D=6$ gauge transformations (\ref{gauge31}).
Parametrizing the six-dimensional gauge parameters as
\bea
\alpha_{\hat\mu\hat\nu}
=
\begin{pmatrix}
\alpha_{\mu\nu} -\eta_{\mu\nu}\,\lambda & 
\frac12\,(\xi_\mu+3\,\Lambda_\mu)
\\
\frac12\,(\xi_\mu+3\,\Lambda_\mu)
& 2\,\lambda 
\end{pmatrix}
\;,
\quad
\beta_{\hat\mu\hat\nu}
=
\begin{pmatrix}
\beta_{\mu\nu} & 
\frac32\,(\xi_\mu-\Lambda_\mu)
\\
\frac32\,(\Lambda_\mu-\xi_\mu)
& 0
\end{pmatrix}
\;,
\eea
their action on the various components of (\ref{CAB31}) is  derived as
\bea
\delta A_\mu &=& \partial_\mu\lambda +\frac12\,\partialsix  \left(\xi_\mu -3\,\Lambda_\mu\right)\;,\nonumber\\
\delta B_{\mu\nu} &=& 2\,\partial_{[\mu}\Lambda_{\nu]} + \frac{1}{3}\,\partialsix \beta_{\mu\nu}\;,\nonumber\\
\delta h_{\mu\nu} &=& 2\, \partial_{(\mu}\xi_{\nu)} +\eta_{\mu\nu}\,\partialsix \lambda - \partialsix  \alpha_{\mu\nu} \;,\nonumber\\
\delta C_{\mu\nu,\rho} &=& 2\,\partial_{[\mu} \alpha_{\nu]\rho} + \partial_\rho \beta_{\mu\nu} 
- \partial_{[\rho} \beta_{\mu\nu]} + \partialsix \left (\xi_{[\mu}\eta_{\nu]\rho} -3\,\Lambda_{[\mu}\eta_{\nu]\rho}\right)\;.
\label{gauge_31}
\eea
With the field $u_{\mu\nu}$ defined by equation (\ref{defc}), its gauge variation is found by integrating up the variation of 
(\ref{defc}) and takes the form
\bea
\delta u_{\mu\nu} &=& \partial_{[\mu}\xi_{\nu]}
+ \frac{1}{6}\,\varepsilon_{\mu\nu\rho\sigma\tau}\partial^\rho \beta^{\sigma\tau}
 + \frac{1}{2}\,\partialsix  \beta_{\mu\nu} 
 \;.
 \label{gaugeu}
\eea

For later use, let us note that contraction of (\ref{defc2}) with the fully antisymmetric $\varepsilon$-tensor
yields
\bea
\frac1{6}\,
\partial^\rho C_{\mu\nu,\rho}
+\frac1{3}\,
\partial_{[\mu} C_{\nu]\rho}{}^{\rho}
+\frac1{6}\, \varepsilon_{\mu\nu\rho\sigma\tau} \partial^{\rho} u^{\sigma\tau}
&=&
\frac12\,  \partialsix  \Big( u_{\mu\nu} 
-\frac32\, B_{\mu\nu} \Big)
\;,
\label{dCvar}
\eea
while contraction gives rise to
\bea
\partial_{\mu} h_{\nu}{}^\mu-\partial_\nu h_\mu{}^\mu
+\partialsix   C_{\mu\nu}{}^\mu
+4\,\partialsix   A_{\nu} 
&=&
2\,\partial^\mu u_{\mu\nu}
\;.
\label{divc2}
\eea
This gives rise to an equivalent rewriting of (\ref{defc2}) as
\bea
\partial_{[\mu} h_{\nu]\rho}
+\partial_{\sigma} h_{[\mu}{}^\sigma  \eta_{\nu]\rho} +\eta_{\rho[\mu} \partial_{\nu]} h_\sigma{}^\sigma  
&\!\!=\!\!&
-\frac1{4}\,\varepsilon_{\mu\nu\kappa\lambda\tau}\,\partial^\kappa 
\left(C^{\lambda\tau}{}_{\rho} +\varepsilon^{\lambda\tau\alpha\beta\sigma}\,u_{\alpha\beta} \eta_{\rho\sigma} \right)
-3\,\partialsix   A_{[\mu} \eta_{\nu]\rho}
\nonumber\\
&&{}
+\frac14\,\varepsilon_{\mu\nu\rho\kappa\lambda}\,  \partialsix  \Big( u^{\kappa\lambda} 
-\frac32\, B^{\kappa\lambda}  \Big)
-\frac1{2}\,\partialsix   C_{\mu\nu,\rho}
-\partialsix   C_{\sigma[\mu}{}^\sigma\eta_{\nu]\rho}
\;.\nonumber \\
\eea
Let us further note that taking the divergence of (\ref{dualFH6})
yields the Maxwell type equation
\bea
\partial^\mu F_{\mu\nu} &=&
 \frac12 \, \partialsix  \partial_{\mu} h_{\nu}{}^\mu
 -\frac12 \, \partialsix \partial_\nu h_\mu{}^\mu
+\frac1{2}\,\partialsix  \partialsix  C_{\mu\nu}{}^\mu
-\frac32\,\partialsix  \partial^\mu B_{\mu\nu}
+2\,\partialsix  \partialsix   A_{\nu}
\;,
\label{Maxwell31}
\eea
where we have used (\ref{divc2}) in order to eliminate the divergence of $u_{\mu\nu}$\,.

For the remaining fields of the ${\cal N}=(3,1)$ model, the 5+1 Kaluza-Klein split is achieved just
as for the ${\cal N}=(2,2)$ model discussed above. The six-dimensional field equations of the
14 vector fields and 28 scalar fields take the form obtained from variation of Lagrangians of the form (\ref{LAP22}),
respectively. The field equations of the 12 selfdual forms take the form
\bea
H_{\mu\nu\rho}{}^a &=& 
\frac12\,\varepsilon_{\mu\nu\rho\kappa\lambda}\,
(F^{\kappa\lambda\,a}+\partialsix  B^{\kappa\lambda\,a})
\;,
\label{sd31KK5}
\eea
after splitting the two-forms according to $\{B_{\hat\mu\hat\nu}{}^a\}=\{B_{\mu\nu}{}^a, B_{\mu 5}{}^a \equiv A_\mu{}^a\}$\,.
The equations may be integrated up to an action in precise analogy with (\ref{LBB22}),
c.f.\ the discussion in appendix~\ref{app:ExFT}.

\subsection{The ${\cal N}=(4,0)$ Model}

In this model, the exotic graviton is given by the rank four tensor (\ref{T40}) whose dynamics is defined by the selfduality
equations (\ref{6DsdT40}) for its second-order curvature. According to the split of coordinates (\ref{split51}), we parametrize
the various components of this field as
\bea
\{ T_{\hat\mu\hat\nu,\hat\rho\hat\sigma} \} &=&
\left\{ T_{\mu\nu,\rho\sigma}\,;\;T_{\mu\nu,\rho5} = C_{\mu\nu,\rho}\,;\; T_{\mu5,\nu5} = h_{\mu\nu} \right\}
\;.
\label{TCh40}
\eea
After dimensional reduction to five dimensions, these fields describe the graviton, dual graviton and double dual graviton, respectively.
Explicitly, in this parametrization the six-dimensional field equations (\ref{6DsdT40})  split into two equations
\bea
R_{\mu\nu,\rho\sigma}
&=&
\frac12\,\partialsix  \partial_{\mu}{C}_{\rho\sigma,\nu}
-\frac12\,\partialsix  \partial_{\nu}{C}_{\rho\sigma,\mu}
+\frac12\, \partialsix \partial_{\rho} {C}_{\mu\nu,\sigma}
-\frac12\,\partialsix  \partial_{\sigma} {C}_{\mu\nu,\rho}
\nonumber\\
&&{} 
+\frac12\,\varepsilon_{\mu\nu\kappa\lambda\tau}\,
\partial_{[\rho} \partial^{\kappa}{C}^{\lambda\tau}{}_{\sigma]}
+\frac14\,\varepsilon_{\mu\nu\kappa\lambda\tau}\,
\partialsix  \partial^{\kappa}T^{\lambda\tau}{}_{\rho\sigma}
+\frac12\, \partialsix  \partialsix T_{\mu\nu,\rho\sigma}
\;,\qquad
\label{6Dsd1}\\[2ex]
\varepsilon_{\mu\nu\alpha\beta\gamma}\,
\partial^\alpha \partial_{[\rho} T_{\sigma\tau]}{}^{\beta\gamma}
&=&
- 2\,\partial_{\mu} \partial_{[\rho}{C}_{\sigma\tau],\nu}
+ 2\,\partial_{\nu} \partial_{[\rho}{C}_{\sigma\tau],\mu}
- 2\,\partialsix  \partial_{[\rho}T_{\sigma\tau],\mu\nu}
\;,
\label{6Dsd2}
\eea
with the linearized Riemann tensor $R_{\mu\nu,\rho\sigma}$ defined as in (\ref{Riemann6D}) 
for the field $h_{\mu\nu}$\,.
The second equation (\ref{6Dsd2}) has the form of a curl in $[\rho\sigma\tau]$ and can be integrated up into
\bea
\frac12\,\varepsilon_{\mu\nu\alpha\beta\gamma}\,
\partial^\alpha  T_{\sigma\tau}{}^{\beta\gamma}
+ \partial_{\mu} {C}_{\sigma\tau,\nu}
- \partial_{\nu} {C}_{\sigma\tau,\mu}
+ \partialsix  T_{\sigma\tau,\mu\nu}
&=&
2\, \partial_{[\sigma} v_{\tau],\mu\nu}
\;,
\label{defv}
\eea
up to a tensor $v_{\tau,\mu\nu}=-v_{\tau,\nu\mu}$, determined by this equation up
to the gauge freedom $\delta v_{\tau,\mu\nu}=\partial_\tau \zeta_{\mu\nu}$\,.
Combining (\ref{defv}) with the first field equation (\ref{6Dsd1}), we find
\bea
R_{\mu\nu,\rho\sigma}
&=&
\frac12\, \partialsix \partial_{\rho} {C}_{\mu\nu,\sigma}
-\frac12\, \partialsix  \partial_{\sigma} {C}_{\mu\nu,\rho}
+\frac12\, \varepsilon_{\mu\nu\kappa\lambda\tau}\,
\partial_{[\rho} \partial^{\kappa} {C}^{\lambda\tau}{}_{\sigma]}
+\partialsix \partial_{[\rho} v_{\sigma],\mu\nu}
\;,
\label{6Dsd1A}
\eea
which in turn is a curl in $[\rho\sigma]$ and can be integrated up into
\bea
\partial_{[\mu} h_{\nu]\rho}
+\frac1{4}\,\varepsilon_{\mu\nu\lambda\sigma\tau}\,
 \partial^{\lambda}{C}^{\sigma\tau}{}_{\rho}
+ \frac1{2}\,\partialsix  {C}_{\mu\nu,\rho}
+\frac12\,\partialsix  v_{\rho,\mu\nu}
&=&
\partial_\rho u_{\mu\nu}
\;,
\label{defu40}
\eea
up to an antisymmetric field $u_{\mu\nu}=-u_{\nu\mu}$\,.
As for the ${\cal N}=(3,1)$ model, we have obtained an equivalent reformulation of the dynamics
in terms of two first-order equations (\ref{defv}) and (\ref{defu40})
from which the original second-order field equations (\ref{6Dsd1}), (\ref{6Dsd2}), can be obtained by derivation.
After reduction to five dimensions, equations (\ref{defv}) and (\ref{defu40}) describe the
duality relations between graviton and dual graviton and between
dual graviton and double dual graviton, respectively.
In particular, equation (\ref{defu40}) differs from equation (\ref{defc2}) in the ${\cal N}=(3,1)$ model
only if fields depend on the sixth coordinate.

It is instructive to work out the gauge symmetries of these equations which originate 
from the $D=6$ gauge transformations (\ref{gauge40}).
Parametrizing the six-dimensional gauge parameters as
\bea
\left\{
\lambda_{\hat\rho,\hat\mu\hat\nu}
\right\} &=&
\left\{
\lambda_{\rho,\mu\nu}\,;\;
\lambda_{\mu,\nu 5}=2\,\alpha_{\mu\nu}-\frac23\,\beta_{\mu\nu}\,;\;
\lambda_{5,\mu5} = 2\,\xi_\mu
\right\} 
\;,
\eea
with symmetric $\alpha_{\mu\nu}$, and antisymmetric $\beta_{\mu\nu}$,
their action on the various components of (\ref{TCh40}) is derived as
\bea
\delta h_{\mu\nu} &=& 
2\,\partial_{(\mu} \xi_{\nu)}
-2\,\partialsix  {\alpha}_{\mu\nu}
\;,\nonumber\\
\delta {C}_{\mu\nu,\rho} &=&
2\,\partial_{[\mu} \alpha_{\nu]\rho}  
+\partial_\rho{\beta}_{\mu\nu}
-\partial_{[\rho}{\beta}_{\mu\nu]}
-\frac12\,\partialsix   \lambda_{\rho,\mu\nu} 
\;,
\nonumber\\
\delta T_{\mu\nu,\rho\sigma} &=& 
\partial_{[\mu} \lambda_{\nu],\rho\sigma} + 
\partial_{[\rho} \lambda_{\sigma],\mu\nu} 
\;.
\label{gauge40_1}
\eea
Gauge variations of the two new fields $v_{\rho,\mu\nu}$ and $u_{\mu\nu}$ are obtained
by integrating up the variation of (\ref{defv}) and (\ref{defu40}), respectively, giving rise to
\bea
\delta u_{\mu\nu}
&=&
 \partial_{[\mu} \xi_{\nu]}
+ \frac1{6}\, \varepsilon_{\mu\nu\lambda\sigma\tau}\partial^{\lambda} \beta^{\sigma\tau} 
+ \frac1{3}\,\partialsix  \beta_{\mu\nu} 
+ \frac1{2}\,\partialsix  \zeta_{\mu\nu} 
\;,
\nonumber\\
 \delta v_{\rho,\mu\nu}
&=&
\frac14\,\varepsilon_{\mu\nu\kappa\lambda\sigma}\,
\partial^\kappa   \lambda_{\rho}{}^{\lambda\sigma}  
+2\,\partial_{[\mu} \alpha_{\nu]\rho} 
+\frac23\,\partial_{[\mu} \beta_{\nu]\rho}
+\partial_\rho\,\zeta_{\mu\nu}
+\frac{1}{2}\, \partialsix   \lambda_{\rho,\mu\nu}
\;,
\label{symmetries40}
\eea
where the antisymmetric gauge parameter $\zeta_{\mu\nu}=-\zeta_{\nu\mu}$ has been introduced after (\ref{defv}).

Let us finally note that from (\ref{6Dsd1}) and (\ref{6Dsd1A}), we may obtain the modified Einstein equations
\bea
\mathring{\cal G}_{\mu\nu} 
&=&
-\frac12\, \partialsix \partial^{\rho} {C}_{\rho(\mu,\nu)}
-\frac12\,\partialsix  \partial_{(\mu} {C}_{\nu)\rho}{}^{\rho}
+\frac12\,\partialsix \partial_{\rho} v_{(\mu,\nu)}{}^\rho
-\frac12\,\partialsix \partial_{(\mu} v^\rho{}_{\nu)\rho}
\nonumber\\
&&{}
+\frac12\,\eta_{\mu\nu}\, \partialsix \partial^{\rho} {C}_{\rho\sigma}{}^\sigma
-\frac12\,\eta_{\mu\nu}\,\partialsix \partial_{\rho} v_{\sigma}{}^{\sigma\rho}
\;,
\label{Einstein40}
\eea
with the linearized Einstein tensor $\mathring{\cal G}_{\mu\nu}$ defined as 
\bea
\mathring{\cal G}_{\mu\nu} &=& 
-\partial^\rho \partial_{(\mu} h_{\nu)\rho}
+\frac12\,\partial_\rho \partial^{\rho} h_{\mu\nu}
+\frac12\,\partial_\mu \partial_{\nu} h_\rho{}^\rho
+\frac12\,\eta_{\mu\nu}\,\partial^\rho \partial^\sigma h_{\rho\sigma}
-\frac12\,\eta_{\mu\nu}\,\partial_\rho \partial^{\rho} h_{\sigma}{}^\sigma
\;,
\label{G40}
\eea
which differs from the previous models by the absence of covariant derivatives, c.f.~(\ref{GcovDh}).

For the remaining fields of the ${\cal N}=(4,0)$ model, the 5+1 Kaluza-Klein split is achieved just
as for the previous models discussed above. The field equations of the 42 scalar fields are 
obtained from variation of a Lagrangian of the form ${\cal L}_\phi$ in (\ref{LAP22}). 
The field equations of the 27 selfdual forms take the form of (\ref{sd31KK5}) above,
again after splitting the two-forms according to 
$\{B_{\hat\mu\hat\nu}{}^a\}=\{B_{\mu\nu}{}^a, B_{\mu 5}{}^a \equiv A_\mu{}^a\}$\,.

\section{Actions for (free) exotic graviton fields}

In the above, we have reformulated the dynamics of the six-dimensional exotic tensor fields in
terms of first order differential equations upon breaking six-dimensional Poincar\'e invariance
according to the split (\ref{split51}), and introducing some additional tensor fields.
As a key property of the resulting equations, we have put the dynamics of the different models
into a form which reduces to the {\em same} equations after dimensional reduction $\partialsix\rightarrow0$.
E.g.\ all three models feature linearized Einstein equations for the field $h_{\mu\nu}$, given by
(\ref{Einstein22}), (\ref{Einstein31}), and (\ref{Einstein40}), respectively. The three equations only differ by terms
carrying explicit derivatives along the sixth dimension.
We will use this as a guiding principle to construct uniform Lagrangians for the ${\cal N}=(3,1)$ and the ${\cal N}=(4,0)$ model
which after setting $\partialsix\rightarrow0$ both reduce to the Lagrangian (\ref{5D22}) of linearized
$D=5$ maximal supergravity.

This construction follows the toy model of $D=6$ selfdual tensor fields whose dynamics can be described by a Lagrangian (\ref{LBB22})
\bea
{\cal L}_{\,{\Yboxdim{4pt}\yng(1,1)}} &=& 
-\frac14\,\left( F_{\mu\nu}{}+ \partialsix  B_{\mu\nu}{} \right)\left( F^{\mu\nu}+ \partialsix  B^{\mu\nu} \right)
- \frac1{24}\,\varepsilon^{\mu\nu\rho\sigma\tau}\, \partialsix  B_{\mu\nu} \,H_{\rho\sigma\tau}
\;,
\label{LSD}
\eea
after a Kaluza-Klein ($5+1$) decomposition $\{B_{\hat\mu\hat\nu}\}=\{B_{\mu\nu}, B_{\mu 5} \equiv A_\mu\}$ of the six-dimensional tensor field.
After dimensional reduction to five dimensions, the 3 degrees of freedom of the selfdual tensor field are
described as a massless vector with the standard Maxwell Lagrangian to which (\ref{LSD}) reduces at $\partialsix\rightarrow0$.
In presence of the sixth dimension, variation of the Lagrangian (\ref{LSD}) w.r.t.\ the vector field gives rise to modified Maxwell equations 
while variation w.r.t.\ the tensor field yields the duality equation relating $A_\mu$ and $B_{\mu\nu}$, which is of first order
in the derivatives $\partial_\mu$ and appears under a global $\partialsix $ derivative. Combining these two equations one may
infer the full six-dimensional selfduality equation. Details are spelled out in appendix~\ref{app:ExFT}.
The Lagrangians for exotic gravitons are constructed in analogy to (\ref{LSD}) with the role of $A_\mu$ and $B_{\mu\nu}$
now taken by the graviton $h_{\mu\nu}$ and its duals, respectively.

\subsection{Action for the ${\cal N}=(3,1)$ Model}

The main result of this subsection is the following: the first order field equations (\ref{dualFH6}) and (\ref{defc2}),
which describe the dynamics of the six-dimensional exotic graviton field $C_{\hat\mu\hat\nu,\hat\rho}$ in the ${\cal N}=(3,1)$ model,
can be derived from the Lagrangian
\begin{equation}
\begin{split}
{\cal L}_{\,{\Yboxdim{4pt} \yng(2,1)}}  \;=\; 
& 
-\frac{1}{4}\,\widehat{\Omega}^{\mu\nu\rho} \widehat{\Omega}_{\mu\nu\rho}
+\frac12\,\widehat{\Omega}^{\mu\nu\rho}\widehat{\Omega}_{\nu\rho\mu}
+\widehat{\Omega}^{\mu}\widehat{\Omega}_{\mu}
-\frac{1}{16}\, \varepsilon^{\mu\nu\rho\sigma\tau} \partialsix  \widehat{C}_{\mu\nu,}{}^{\lambda}  
\partial_{\rho} \widehat{C}_{\sigma\tau,\lambda} 
\\& 
-\frac{3}{4}\,{\cal F}^{\mu\nu}{\cal F}_{\mu\nu}
-\frac{9}{16} \, \varepsilon^{\mu\nu\rho\sigma\tau} \partialsix  B_{\mu\nu} \partial_{\rho}B_{\sigma\tau} 
-\frac{3}{16}\, \varepsilon^{\mu\nu\rho\sigma\tau}\,
\partialsix  B_{\mu\nu} \,\partialsix  \widehat{C}_{\rho\sigma,\tau}
\;,
\end{split}
\label{L31}
\end{equation}
with
\be
\begin{split}
\widehat{\Omega}_{\mu\nu\rho}&\equiv \partial_{[\mu} h_{\nu]\rho} 
-\partialsix  A_{[\mu} \eta_{\nu]\rho} +\frac{1}{2}\, \partialsix  \widehat{C}_{\mu\nu,\rho}
\;,\\
\widehat{C}_{\mu\nu,\rho} &\equiv
C_{\mu\nu,\rho} + \varepsilon_{\mu\nu\rho\sigma\tau}\,u^{\sigma\tau}
\;, \\
{\cal F}_{\mu\nu} &\equiv 2\, \partial_{[\mu}A_{\nu]} 
+\frac32 \, \partialsix  B_{\mu\nu} 
\;.
\end{split}
\label{OC}
\ee 
The Lagrangian (\ref{L31}) is invariant under the gauge transformations (\ref{gauge_31}), (\ref{gaugeu}).
After reduction to five dimensions, i.e.\ at $\partialsix\rightarrow0$, this Lagrangian reduces
to the Fierz-Pauli Lagrangian for $h_{\mu\nu}$ together with a free Maxwell Lagrangian for $A_\mu$;
the dual fields $\widehat{C}_{\mu\nu,\rho}$ and $B_{\mu\nu}$ drop out in this limit.
In presence of the sixth dimension, variation of the Lagrangian (\ref{L31}) w.r.t.\ to the dual fields
yields the first-order duality equations (\ref{dualFH6}) and (\ref{defc2}), however under an overall
derivative $\partialsix$. Together, one recovers the full six-dimensional dynamics. 
Details of the equivalence are presented in appendix~\ref{app:31}.

The bosonic Lagrangian for the full ${\cal N}=(3,1)$ model is then given by combining (\ref{L31}) with the Lagrangians of the type
(\ref{LAP22}) and (\ref{LSD}) for the remaining matter fields of the theory. Putting everything together, we obtain
\bea
{\cal L}_{(3,1)} &=& 
{\cal L}_{\,{\Yboxdim{4pt} \yng(2,1)}} \;
 - \frac{1}{2}\left(\partial^\mu \phi^i -\partialsix  A^{\mu\,i} \right)
 \left(\partial_\mu \phi^i -\partialsix  A_\mu{}^{i} \right)
- \frac{1}{2}\,\partial^\mu \phi^\alpha \partial_\mu \phi^\alpha- \frac{1}{2}\,\partialsix  \phi^\alpha \partialsix  \phi^\alpha
\label{L31full}\\
&&{}
 -\frac{1}{4}\, F^{\mu\nu\,i}F_{\mu\nu}{}^i 
-\frac14\,\left( F_{\mu\nu}{}^a+ \partialsix  B_{\mu\nu}{}^a \right)\left( F^{\mu\nu\,a}+ \partialsix  B^{\mu\nu\,a} \right)
- \frac1{24}\,\varepsilon^{\mu\nu\rho\sigma\tau}\, \partialsix  B_{\mu\nu}{}^a \,H_{\rho\sigma\tau}{}^a
\;,
\nonumber
\eea
with indices ranging along
\bea
i=1, \dots, 14\;,\quad
\alpha=1, \dots, 28\;,\quad
a=1, \dots, 12
\;.
\eea
After dimensional reduction to five dimensions (and rescaling of the vector field $A_\mu$), 
this Lagrangian coincides with the Lagrangian (\ref{5D22}) of linearized maximal supergravity.
The Lagrangian (\ref{L31full}) describes the full six-dimensional theory, with the field content
of five-dimensional maximal supergravity enhanced by the field $\widehat{C}_{\mu\nu,\rho}$\,.
$D=6$ Poincar\'e invariance is no longer manifest although it can still be realized on the equations of motion.

\subsection{Action for the ${\cal N}=(4,0)$ Model}

The main result of this subsection is the following:  the first order field equations (\ref{defv}) and (\ref{defu40}),
which describe the dynamics of the six-dimensional exotic graviton field $T_{\hat\mu\hat\nu,\hat\rho\hat\sigma}$
 in the ${\cal N}=(4,0)$ model, can be derived from the Lagrangian
 \bea
{\cal L}_{\,{\Yboxdim{4pt}\yng(2,2)}}&=&
-\frac14\,\widehat{\Omega}^{\mu\nu\rho}\widehat{\Omega}_{\mu\nu\rho}
+\frac12\,\widehat{\Omega}^{\mu\nu\rho} \widehat{\Omega}_{\nu\rho\mu}
+\widehat{\Omega}^{\mu}\widehat{\Omega}_{\mu}
- \frac1{8}\,\varepsilon_{\mu\nu\sigma \kappa\lambda}\,
\partial^{\mu} \widehat{{C}}{}^{\nu\sigma}{}_{\rho} \,\partialsix  \widehat{{C}}{}^{\kappa\lambda,\rho}
\nonumber\\
&&{}
+ \frac1{32}\,\varepsilon_{\mu\nu\sigma \kappa\lambda}\,
\partial^{\mu} {\cal C}{}^{\nu\sigma}{}_{\rho} \,\partialsix  {\cal C}{}^{\kappa\lambda,\rho}
- \frac1{8}\,\partialsix  {\cal C}{}_{\sigma\tau,\nu}\,\partial_{\mu} T^{\mu\nu,\sigma\tau}
+\frac1{4}\,\partialsix  {\cal C}{}_{\kappa\lambda,\tau}\,\partial^{\kappa}  T^{\lambda\sigma,\tau}{}_{\sigma}
\nonumber\\
&&{}
+ \frac1{4}\,\partial_{\nu} {\cal C}{}_{\sigma\mu}{}^{\mu}\,\partialsix  T^{\sigma\tau,\nu}{}_{\tau}
- \frac1{8}\,\partialsix  {\cal C}{}_{\sigma\mu}{}^{\mu}\,\partial^{\sigma} T_{\tau\nu}{}^{\tau\nu}
-\frac1{64}\,\varepsilon_{\mu\nu\alpha\beta\gamma}\,
\partial^\alpha  T_{\sigma\tau}{}^{\beta\gamma}
\,\partialsix  T^{\mu\nu,\sigma\tau}
\nonumber\\
&&{}
- \frac1{32}\,\partialsix  T_{\sigma\tau,\mu\nu}\,\partialsix  T^{\mu\nu,\sigma\tau}
+\frac18\,\partialsix  T_{\sigma\mu,\nu}{}^\mu\,\partialsix  T^{\sigma\tau,\nu}{}_{\tau}
- \frac1{32}\,\partialsix  T_{\mu\nu}{}^{\mu\nu} \,\partialsix  T_{\sigma\tau}{}^{\sigma\tau}
\;,
 \label{L40}
\eea
with
\bea
  \widehat{\Omega}_{\mu\nu\rho}
  &=&
  \partial_{[\mu} h_{\nu]\rho}  
 +\partialsix  \widehat{{C}}_{\mu\nu,\rho}
 -\frac{1}{2}\partialsix  {\cal{C}}_{\mu\nu,\rho}
 \;,\nonumber\\
 \widehat{{C}}_{\mu\nu,\rho} &=& 
 {{C}}_{\mu\nu,\rho} + \varepsilon_{\mu\nu\rho\sigma\tau}\,u^{\sigma\tau}
 \;,
 \nonumber\\
  {\cal C}_{\mu\nu,\rho} &=& 
 {{C}}_{\mu\nu,\rho} - v_{\rho,\mu\nu} + 
 2 \,v_{[\rho,\mu\nu]} 
 + 2\,\varepsilon_{\mu\nu\rho\sigma\tau}\,u^{\sigma\tau}
 \;.
\label{OCC} 
\eea
After reduction to five dimensions, i.e.\ at $\partialsix\rightarrow0$, this Lagrangian reduces
to the Fierz-Pauli Lagrangian for $h_{\mu\nu}$;
the dual fields $\widehat{C}_{\mu\nu,\rho}$, ${\cal C}_{\mu\nu,\rho}$, and $T_{\mu\nu,\rho\sigma}$ drop out in this limit.
In presence of the sixth dimension, variation of the Lagrangian (\ref{L40}) w.r.t.\ to the dual fields
yields the first-order duality equations (\ref{defv}) and (\ref{defu40}), however under an overall
derivative $\partialsix$. Together, one recovers the full six-dimensional dynamics. 
The computation works in close analogy with the derivation for the ${\cal N}=(3,1)$ model, c.f.~appendix~\ref{app:31}.

Let us spell out the gauge transformations (\ref{gauge40_1}), (\ref{symmetries40}) in terms of the fields (\ref{OCC})
\begin{align}
\delta \widehat{\Omega}_{\mu\nu\rho}&=
\partial_\rho\partial_{[\mu}\xi_{\nu]}
-\frac23\,\partialsix\partial_{[\mu}{\beta}_{\nu]\rho}
-\partialsix\partial_{[\mu}{\zeta}_{\nu]\rho}
+\frac{1}{4}\,\partial_y \partial^{\kappa}  {\lambda}_{[\mu}{}^{\sigma\tau}\,\varepsilon_{\nu]\kappa\sigma\tau\rho}
\,,\nonumber\\[1ex]
\delta \widehat{C}_{\mu\nu,\rho}&=
2\,\partial_{[\mu}{\alpha}_{\nu]\rho}
-2\,\partial_{[\mu}{\beta}_{\nu]\rho}
+\varepsilon_{\mu\nu\rho\sigma\tau}\,\partial^\sigma \xi^\tau
+\frac{1}{3}\,\varepsilon_{\mu\nu\rho\sigma\tau}\partialsix \Big({\beta}^{\sigma\tau}+\frac32\, {\zeta}^{\sigma\tau}\Big)
-\frac{1}{2}\partialsix{\lambda}_{\rho,\mu\nu}
\,,\nonumber\\[1ex]
\delta{\cal C}_{\mu\nu,\rho}&=
2\,\varepsilon_{\mu\nu\rho\sigma\tau}\,\partial^\sigma\xi^\tau
-\frac83\,\partial_{[\mu}{\beta}_{\nu]\rho}
+2\,\partial_{[\mu}\zeta_{\nu]\rho}
+\frac23\,\varepsilon_{\mu\nu\rho\sigma\tau}\partialsix \Big({\beta}^{\sigma\tau}+\frac32\, {\zeta}^{\sigma\tau}\Big)
\nonumber\\
&\hspace{1cm}
+\frac{1}{2}\,\varepsilon_{\kappa\sigma\tau\rho[\mu}\,\partial^{\kappa} {\lambda}_{\nu]}{}^{\sigma\tau}
-\partialsix {\lambda}_{\rho,\mu\nu}
\;,
\label{gauge40OCC}
\end{align}
which allows to confirm gauge invariance of the Lagrangian (\ref{L40}).

The bosonic Lagrangian for the full ${\cal N}=(4,0)$ model is finally given by combining (\ref{L40}) with the Lagrangians of the type 
(\ref{LAP22}) and (\ref{LSD}) for the remaining matter fields of the theory. Putting everything together, we obtain
\bea
{\cal L}_{(4,0)} &=& 
{\cal L}_{\,{\Yboxdim{4pt} \yng(2,2)}} \;
-\frac14\,\left( F_{\mu\nu}{}^M+ \partialsix  B_{\mu\nu}{}^M \right)\left( F^{\mu\nu\,M}+ \partialsix  B^{\mu\nu\,M} \right)
- \frac1{24}\,\varepsilon^{\mu\nu\rho\sigma\tau}\, \partialsix  B_{\mu\nu}{}^M \,H_{\rho\sigma\tau}{}^M
\nonumber\\
&&{}
- \frac{1}{2}\,\partial^\mu \phi^A \partial_\mu \phi^A- \frac{1}{2}\,\partialsix  \phi^A \partialsix  \phi^A
\;,
\label{L40full}
\eea
with indices ranging along
\bea
M=1, \dots, 27\;,\quad
A=1, \dots, 42
\;.
\eea
After dimensional reduction to five dimensions, 
this Lagrangian coincides with the Lagrangian (\ref{5D22}) of linearized maximal supergravity.
The Lagrangian (\ref{L40full}) describes the full six-dimensional theory, with the field content
of five-dimensional maximal supergravity enhanced by the fields $\widehat{C}_{\mu\nu,\rho}$,
${\cal C}_{\mu\nu,\rho}$, and $T_{\mu\nu,\rho\sigma}$\,.
$D=6$ Poincar\'e invariance is no longer manifest although it can still be realized on the equations of motion.

\section{Progress toward an exceptional master action} 

In the previous sections, we have constructed Lagrangians (\ref{L22h0}), (\ref{L31full}), and (\ref{L40full}),
for the three six-dimensional models which share
a number of universal features and structures.
In particular, after dimensional reduction to five dimensions they all reduce to the same Lagrangian (\ref{5D22})
corresponding to linearized maximal supergravity in five dimensions.
The three distinct six-dimensional theories are then described as different extensions of this Lagrangian by
terms carrying derivatives along the sixth dimension. In the various matter sectors, these terms ensure covariantization under
non-trivial gauge structures and provide sources to the field equations of five-dimensional supergravity.

This reformulation within a common framework is very much in the spirit of exceptional field theories.
In that framework, higher-dimensional supergravity theories are reformulated in terms of the field content of a 
lower-dimensional supergravity keeping the dependence on all coordinates.
More precisely, their formulation is based on a split of coordinates into $D$ external and $n$ internal coordinates
of which the latter are formally embedded into a fundamental representation ${\cal R}_v$ of the global symmetry group
${\rm E}_{11-D,(11-D)}$ of $D$-dimensional maximal supergravity. Different embeddings of the internal coordinates
into ${\cal R}_v$ then correspond to different higher-dimensional origins.
Here, we will discuss a similar uniform description of the six-dimensional models based on $D=5$ external dimensions 
which encompasses the three different models upon proper identification of the sixth coordinate within the internal coordinates.
As discussed in the introduction this will require an enhancement of the internal coordinates of exceptional field theory
by an additional exotic coordinate related to the singlet central charge in the $D=5$ supersymmetry algebra.

\subsection{Linearized ExFT and embedding of the ${\cal N}=(2,2)$ model}

The theory relevant for our discussion is ${\rm E}_{6(6)}$ exceptional field theory (ExFT) \cite{Hohm:2013pua,Hohm:2013vpa}.
Its bosonic field content is given by a graviton $g_{\mu\nu}$ together with 27 vector fields ${\cal A}_\mu{}^M$
and their dual tensors ${\cal B}_{\mu\nu\,M}$, together with 42 scalars parametrizing the internal metric 
${\cal M}_{MN}=({\cal V}{\cal V}^T)_{MN}$ with ${\cal V}$ a representative of the coset space ${\rm E}_{6(6)}/{\rm USp}(8)$.
Fields depend on 5 external and 27 internal coordinates with the latter transforming in the fundamental 
${\bf 27}$ of ${\rm E}_{6(6)}$ and with internal coordinate dependence of the fields
restricted by the section constraint \cite{Coimbra:2011ky}
\bea
d^{KMN}\,\partial_M \otimes \partial_N &=& 0
\;,
\label{section6}
\eea
with the two differential operators acting on any couple of fields and gauge parameters of the theory.
The tensor $d^{KMN}$ denotes the cubic totally symmetric E$_{6(6)}$ invariant tensor, which we normalize
as $d^{MNP}d_{MNQ}=\delta_Q{}^P$\,.
The section condition (\ref{section6}) admits two inequivalent solutions \cite{Hohm:2013pua} which reduce the 
internal coordinate dependence of all fields to the 6 internal coordinates from $D=11$ supergravity, or
5 internal coordinates from IIB supergravity, respectively.
For details of the ExFT Lagrangian we refer to \cite{Hohm:2013pua,Hohm:2013vpa}.
Here, we spell out its `free' limit, obtained by linearizing the full theory according to
\bea
g_{\mu\nu} = \eta_{\mu\nu}+h_{\mu\nu}\;,\qquad
{\cal M}_{MN} = \Delta_{MN} + \phi_{MN}
\;,
\label{linear}
\eea
around the constant background given by the Minkowski metric $\eta_{\mu\nu}$
and the identity matrix $\Delta_{MN}$. The scalar fluctuations $\phi_{MN}$ are further constrained
by the coset properties of ${\cal M}_{MN}$\,.
To quadratic order in the fluctuations, the ExFT Lagrangian then yields
\begin{align}
{\cal L}_{\rm ExFT, free}=&   -\frac{1}{4}\,{\Omega}^{\mu\nu\rho} {\Omega}_{\mu\nu\rho}
+\frac12\,{\Omega}^{\mu\nu\rho}{\Omega}_{\nu\rho\mu}
+{\Omega}^{\mu}{\Omega}_{\mu}
- \frac{1}{4}\, {\cal F}^{\mu\nu\,M}\,{\cal F}_{\mu\nu}{}^N\,\Delta_{MN} 
\nonumber\\
&
-\frac{5}{4}\,\sqrt{10}\,\varepsilon^{\mu\nu\rho\sigma\tau}\,
d^{MNK}  \partial_\mu {\cal B}_{\nu\rho\,M} \partial_N {\cal B}_{\sigma\tau\,K}
-\frac1{24}\,D_\mu \phi^{MN}\,D^\mu \phi_{MN}
+{\cal L}_{\rm pot}
\;,
\label{ExFTfree}
\end{align}
with 
indices $M, N$ raised and lowered by $\Delta_{MN}$ and its inverse, 
and with the various elements of (\ref{ExFTfree})
given by
\bea
{\Omega}_{\mu\nu\rho}&=& \partial_{[\mu} h_{\nu]\rho} - \frac23\,\partial_M  {\cal A}_{[\mu}{}^M \eta_{\nu]\rho} 
\;,
\qquad
\Omega_\mu ~\equiv~ \Omega_{\mu\nu}{}^\nu
\;,
\nonumber\\
{\cal F}_{\mu\nu}{}^M &=& 2\,\partial_{[\mu} {\cal A}_{\nu]}{}^M
+ 10\,d^{MNK}\,\partial_N {\cal B}_{\mu\nu\,K}
\;,
\nonumber\\
D_\mu \phi^{MN}
&=& 
\partial_\mu \phi^{MN}
+2\,\partial_K {\cal A}_\mu{}^{(M} {\Delta}^{N)K}
+\frac23\,\partial_K {\cal A}_{\mu}{}^K \Delta^{MN}
-20\,\partial_K {\cal A}_\mu{}^L\,d_{PLR} d^{RK(M} \Delta^{N)P}
\;,
\nonumber\\
  {\cal L}_{\rm pot} &=&
  -\frac{1}{24}{\Delta}^{MN}\partial_M{\phi}^{KL}\,\partial_N{\phi}_{KL}
  +\frac{1}{2} {\Delta}^{MN}\partial_M{\phi}^{KL}\partial_L{\phi}_{NK}
  -\frac{1}{2}\,\partial_M h_\nu{}^\nu \,\partial_N{\phi}^{MN}
  \nonumber\\
  &&{}
  +\frac{1}{4}\,  {\Delta}^{MN}\,\partial_M h_\mu{}^\mu\,\partial_N h_\nu{}^\nu
  -\frac{1}{4}\,{\Delta}^{MN}\,\partial_M h^{\mu\nu}\partial_N h_{\mu\nu}\;. 
\label{ExFTobjects}
\eea

The Lagrangian we have presented above for the six-dimensional ${\cal N}=(2,2)$ model
naturally fits into this framework. This does not come as a surprise since the six-dimensional model is nothing but linearized
maximal supergravity known to be described by ${\rm E}_{6(6)}$ ExFT 
upon proper selection of the sixth coordinate among the internal $\partial_M$\,.
This choice is uniquely fixed by the requirement that the resulting theory exhibits the global 
${\rm SO}(5,5)$ symmetry group of maximal six-dimensional supergravity, thus breaking 
\bea
{\rm E}_{6(6)} ~\longrightarrow~ {\rm SO}(5,5)\;,\qquad\quad
{\bf 27} &\longrightarrow& 1 \oplus 16 \oplus 10 \;,\nonumber\\
\{\partial_M\}  &\longrightarrow&  \{\partial_0, \partial_i, \partial_a\}
\;,
\label{E6toSO55}
\eea
and keeping only coordinate-dependence along the ${\rm SO}(5,5)$ singlet.
In this split, the E$_{6(6)}$ invariant symmetric tensor $d^{MNK}$ has the non-vanishing components
\bea
d^{0ab} &=& \frac1{\sqrt{10}}\, \eta^{ab}
\;,\qquad
d^{aij} ~=~\frac{1}{2\,\sqrt{5}}\, (\Gamma^a)^{ij}
\;,
\label{dSO55}
\eea
in terms of ${\rm SO}(5,5)$ $\Gamma$-matrices and its invariant tensor $\eta^{ab}$ of signature $(5,5)$,
showing that the section constraint (\ref{section6}) is trivially satisfied is $\partial_i=0=\partial_a$\,.
Putting this together with the linearized ExFT Lagrangian (\ref{ExFTfree}), and splitting fields as
\bea
\{ {\cal A}_\mu{}^M \} &=& \{ A_\mu, A_\mu{}^i, A_\mu{}^a \}\;,\qquad
\mbox{etc.}\;,
\eea
we arrive at
\begin{align}
{\cal L}_{(2,2)}=&   -\frac{1}{4}\,{\Omega}^{\mu\nu\rho} {\Omega}_{\mu\nu\rho}
+\frac12\,{\Omega}^{\mu\nu\rho}{\Omega}_{\nu\rho\mu}
+{\Omega}^{\mu}{\Omega}_{\mu}
- \frac{1}{4}\, F^{\mu\nu}F_{\mu\nu} 
-\frac{1}{4}\, F^{\mu\nu\,i}F_{\mu\nu}{}^i 
\nonumber\\
&  
-\frac14\,\left( F_{\mu\nu}{}^a+ \partialsix  B_{\mu\nu}{}^a \right)\left( F^{\mu\nu\,a}+ \partialsix  B^{\mu\nu\,a} \right)
-
\frac1{24}\,\varepsilon^{\mu\nu\rho\sigma\tau}\,\eta_{ab}\, \partialsix  B_{\mu\nu}{}^a \,
H_{\rho\sigma\tau}{}^b - \frac{1}{2}\,\partial^\mu \phi^\alpha \partial_\mu \phi^\alpha
\nonumber\\
&
-  \frac{1}{2}\, ( \partial^\mu \phi -\sqrt{\frac83}\,\partialsix  A^\mu) ( \partial_\mu \phi  -\sqrt{\frac83}\,\partialsix  A_\mu  ) 
 - \frac{1}{2} \left(\partial^\mu \phi^i -\partialsix  A^{\mu\,i} \right)
 \left(\partial_\mu \phi^i -\partialsix  A_\mu{}^{i} \right)\;,
 \nonumber\\
&- \frac{1}{2}\,\partialsix  \phi^\alpha \partialsix  \phi^\alpha
+ \frac{5}{6}\, \partialsix  \phi \partialsix  \phi
 - \frac{2}{3}\, \partialsix  h_\sigma{}^\sigma  \partialsix  \phi 
+ \frac{1}{4}\,\partialsix  h_\sigma{}^\sigma \partialsix  h_\rho{}^\rho 
- \frac14\, \partialsix h^{\mu\nu}\partialsix  h_{\mu\nu}
\;,
\label{L22inExFT}
\end{align}
which precisely produces the sum of Lagrangians (\ref{L22h0}), (\ref{LAP22}), (\ref{LBB22}),
after proper rescaling of the singlet scalar field $\phi$\,.
The non-trivial checks of this coincidence include all the coefficients in the various connection terms,
as well as in the St\"uckelberg-type couplings between vector and tensor fields, and the coefficients
in front of the various $\partial_y\phi \partial_y\phi$ terms in the last line.
Again, this is not a surprise but a consequence of the proven equivalence of ExFT with higher-dimensional maximal supergravity.
Note that although the free theory only exhibits a compact ${\rm USp}(4)\times {\rm USp}(4)$ global symmetry,
the couplings exhibited in (\ref{L22inExFT}) are far more constrained than allowed by this symmetry and witness
the underlying ${\rm E}_{6(6)}$ structure broken to ${\rm SO}(5,5)$ according to (\ref{E6toSO55}), (\ref{dSO55}).

The ExFT Lagrangian is to a large extent determined by invariance under generalized internal diffeomorphisms acting with 
a gauge parameter $\Lambda^M$ in the ${\bf 27}$. After linearization (\ref{linear}) these diffeomorphisms act as
\be
\begin{split}
\delta {\phi}_{MN} &=
2\, \Delta_{K(M} \,\partial_{N)} \Lambda^K
+\frac23\,\partial_K \Lambda^K \,\Delta_{MN}
-20 \,d^{PKR}d_{RL(M} \,\Delta_{N)P}\,\partial_K \Lambda^L
\;,\\
\delta {\cal A}_{\mu}{}^M &= \partial_\mu \Lambda^M
\;,\qquad
\delta h_{\mu\nu} = 
 \frac23\, \partial_M\Lambda^M\, \eta_{\mu\nu} 
\;,
\end{split}
\label{gen_diff_lin}
\ee
and one can show invariance of the linearized Lagrangian (\ref{L22inExFT}), 
provided the section constraint (\ref{section6}) is satisfied.

\subsection{Beyond standard ExFT: embedding of the ${\cal N}=(3,1)$ and $(4,0)$ couplings}

As we have discussed in the introduction, the charges carried by the massive BPS multiplets in the reduction
of the ${\cal N}=(3,1)$ and the ${\cal N}=(4,0)$ model, respectively, suggest that an inclusion of these models into the framework of ExFT
necessitates an extension of the space of 27 internal coordinates by an additional exotic coordinate corresponding
to the singlet central charge~\cite{Hull:2000cf}. Denoting derivatives along this coordinate by $\partial_\bullet$, 
this would amount to a relaxation of the standard section constraint
(\ref{section6}) to a constraint of the form
\bea
d^{KMN}\,\partial_M \otimes \partial_N -
\frac1{\sqrt{10}}\,{\Delta}^{KM}\left(\partial_M \otimes \partial_\bullet +\partial_\bullet \otimes \partial_M  \right)
&=& 0
\;,
\label{sectionext}
\eea
which at the present stage only makes sense in the linearized theory where ${\Delta}^{KM}$ is a constant background tensor.
Apart from the standard ExFT solutions
\bea
d^{KMN}\,\partial_M \otimes \partial_N = 0\;,\qquad \partial_\bullet=0
\;,
\eea
of this constraint, which allow the embedding of the ${\cal N}=(2,2)$ model as described above,
the extended section constraint also allows for two exotic solutions
\bea
(3,1) &:&
\partialsix^{(3,1)} ~=~ \frac{2}{\sqrt{3}}\,\partial_0 ~=~ -2\,\partial_\bullet
\;,\qquad
\mbox{with the F$_{4(4)}$ singlet } \partial_0 \subset \partial_M
\;,
\nonumber\\
(4,0) &:&
\partialsix^{(4,0)} ~=~  -\partial_\bullet\;,\quad \partial_M = 0
\;,
\label{section_ext_sol}
\eea
corresponding to the two exotic six-dimensional models
in precise correspondence with the central charges carried by the corresponding BPS multiplets \cite{Hull:2000cf}.
While the (4,0) solution trivially solves the constraint (\ref{sectionext}), the ${\cal N}=(3,1)$ solution is based on the decomposition
\bea
{\rm E}_{6(6)} ~\longrightarrow~ {\rm F}_{4(4)}\;,\qquad\quad
{\bf 27} &\longrightarrow& 1 \oplus 26\;,\nonumber\\
\{\partial_M\}  &\longrightarrow&  \{\partial_0, \partial_A\}
\;,
\label{E6toF4}
\eea
under which the symmetric $d$-tensor decomposes into
\bea
&&
d^{000}=-\frac{2}{\sqrt{30}}\;,\quad
d^{0AB}=\frac{1}{\sqrt{30}}\,\eta^{AB}\;,\quad
d^{ABC}\;,
\label{dF4}
\eea
with the ${\rm F}_{4(4)}$ invariant symmetric tensor $\eta^{AB}$ of signature $(14,12)$, and the 
symmetric invariant tensor $d^{ABC}$ satisfying
\bea
d^{ABC}\eta_{BC}=0\;,\qquad
d_{ABC} d^{ABD}  = \frac{14}{15}\,\delta_C{}^D 
\;.
\eea
This shows explicitly how the (3,1) assignment of (\ref{section_ext_sol}) also provides a solution to 
the extended section constraint (\ref{sectionext}).

It is intriguing to study the fate of diffeomorphism invariance of the ExFT Lagrangian (\ref{ExFTfree})
if the original section constraint is relaxed to (\ref{sectionext}). 
Except for the last term in (\ref{ExFTfree}), the Lagrangian remains manifestly invariant without 
any use of the section constraint. Explicit variation of the potential term 
${\cal L}_{\rm pot}$ 
under linearized diffeomorphisms (\ref{gen_diff_lin}) on the other hand yields (up to total derivatives)
\bea
\delta_\Lambda {\cal L}_{\rm pot} &=&
\left(
5\,\Delta_{LS}\,d^{LMN}d^{KPQ}
- 10 {\Delta}^{MN} \Delta^{KL} \,d_{LSR}d^{RPQ} \right)  \Lambda^S \,\partial_P \partial_Q \partial_M\phi_{NK}
\nonumber\\
&&{}
-10\,h_\mu{}^\mu\,\Delta^{MK}\,d_{KLR} d^{RPQ}\,\partial_M \partial_P \partial_Q\Lambda^L
\;,
\eea
which consistently vanishes modulo the standard section constraint (\ref{section6}).
For the weaker constraint (\ref{sectionext}), this variation no longer vanishes and may be recast in the following form
\bea
\delta_\Lambda {\cal L}_{\rm pot} &=&
\Delta^{KM}   \,  \Lambda^N \,\partial_\bullet \partial_\bullet\partial_M{\phi}_{NK}
-4\, h_\nu{}^\nu \,\partial_\bullet \partial_\bullet \partial_N \Lambda^N \;,
\label{off_pot}
\eea
after repeated use of (\ref{sectionext}) and further manipulation of the expressions.
In order to compensate for this variation
let us first note that there is no possible covariant extension
of the transformation rules (\ref{sectionext}) by terms carrying $\partial_\bullet\Lambda^M$,
such that invariance can only be restored by extending the potential.
A possible such extension is given by
\bea
{\cal L}_{{\rm pot},\bullet}  &=&
 {\cal L}_{\rm pot} 
 -\frac{1}{24}\,\partial_\bullet {\phi}_{MN} \partial_\bullet {\phi}^{MN}
  -\frac{3}{4} \,\partial_\bullet h_\sigma{}^\sigma \partial_\bullet h_\rho{}^\rho  
+\frac{3}{4}\, \partial_\bullet h^{\mu\nu}\partial_\bullet h_{\mu\nu}
\;,
\label{pot_ext}
\eea
and it is straightforward to verify that the variation of the additional terms in (\ref{pot_ext})
precisely cancels the contributions in (\ref{off_pot}), such that
\bea
\delta_\Lambda {\cal L}_{{\rm pot},\bullet}  &=& 0\;.
\eea
For the exotic solutions of the section constraint, the $\partial_\bullet {\phi}_{MN} \partial_\bullet {\phi}^{MN}$ 
terms in (\ref{pot_ext}) give rise
to additional contributions of the type $\partial_y \phi \partial_y \phi$ in the Lagrangian.
Collecting all such terms in (\ref{pot_ext}) for the two exotic solutions (\ref{section_ext_sol}) yields
\bea
{(3,1)} &\longrightarrow& 
- \frac{1}{2}\,\partialsix  \phi^\alpha \partialsix  \phi^\alpha\;,\qquad \alpha=1, \dots, 28
\;,\nonumber\\
(4,0) &\longrightarrow& 
- \frac{1}{2}\,\partialsix  \phi^A \partialsix  \phi^A
\;,\qquad A=1, \dots, 42
\;.
\eea
These are precisely the terms found in our explicit construction of actions (\ref{L31full}) and (\ref{L40full})
above! In other words, the relaxation (\ref{sectionext}) of the section constraint together with generalized diffeomorphism
invariance precisely implies the correct scalar couplings in the Lagrangians of the exotic models.
In addition, the $\partial_\bullet h \partial_\bullet h$ terms in (\ref{pot_ext}) cancel the corresponding terms in 
${\cal L}_{\rm pot}$ (\ref{ExFTobjects}) upon selecting the (3,1) solution of the section constraint (\ref{section_ext_sol}), just as required 
in order to reproduce the correct Lagrangian of the ${\cal N}=(3,1)$ model (\ref{L31}).\footnote{
In contrast, these terms appear in conflict with embedding the spin-2 sector of the ${\cal N}=(4,0)$ model
as they survive under the (4,0) solution in (\ref{section_ext_sol}) but should be absent in the final Lagrangian (\ref{L40}).
We come back to this in section~\ref{subsec:spin2}.
}

We may continue the symmetry analysis for the tensor gauge transformations given by a gauge parameter $\Lambda_{\mu\,M}$
in standard ExFT. For these transformations there is a natural extension of the standard ExFT transformation rules
in presence of the exotic coordinate and exotic fields as
\be
\begin{split}
\delta_{\Lambda_\mu}  A_{\mu}{}^M &= 
-10\,d^{MNK}\,\partial_N \Lambda_{\mu\,K}
- \sqrt{10}\,\Delta^{MK}\,\partial_\bullet \Lambda_{\mu\,K}\;, \\
\delta_{\Lambda_\mu}  B_{\mu\nu\,M} &= 2\, \partial_{[\mu}\Lambda_{\nu]\,M}
\;.
\end{split}
\label{Lambda_mu}
\ee
Computing the action of these transformations on the connection featuring in the
covariant scalar derivatives $D_\mu \phi^{MN}$ in (\ref{ExFTobjects}), we obtain after some manipulation\footnote{
A useful identity for this computation is given by
\bea
d^{PLQ} d_{PSR}d^{KMR}  \partial_K \partial_L
 =
\frac1{10}\,
\delta_S^{K} d^{LQM} \partial_K \partial_L 
+\frac1{20}\,
\delta_S^{M} d^{QKL} \partial_K \partial_L 
+\frac1{20}\,
\delta_S^{Q} d^{MKL} \partial_K \partial_L 
 -\frac12\, d^{QMR}   d_{RSP} d^{PKL} \partial_K \partial_L
 \;,
 \nonumber
\eea
generalizing equations  (2.12), (2.13) of \cite{Hohm:2013vpa}.
}
\bea
\delta_{\Lambda_\mu} D_\mu {\phi}^{MN} &=&
10 \left(
\frac13\,\Delta^{MN}\,\delta_P{}^Q 
+\Delta^{Q(M} \, \delta_P{}^{N)} 
-10\, \Delta^{S(M}  d^{N)QR}   d_{RSP} 
\right) 
d^{PKL} \,\partial_K \partial_L \Lambda_{\mu\,Q}
\nonumber\\
&&{}
- 2\,\sqrt{10} \left(
\frac13\,\Delta^{MN}\, \delta_P{}^{Q}
+  \delta_P{}^{(M}  \Delta^{N)Q} 
-10\,\Delta^{L(M} d^{N)QR}d_{RPL}  \right)
\Delta^{PK} \partial_K \partial_\bullet \Lambda_{\mu\,Q}  
\;.
\nonumber\\
\eea
The resulting expression precisely vanishes with the modified section constraint (\ref{sectionext}).
This shows the necessity of the $\partial_\bullet\Lambda_{\mu\,M}$ terms in (\ref{Lambda_mu})
in order to maintain gauge invariance of the kinetic term $D_\mu \phi^{MN} D^\mu \phi_{MN}$
in presence of the relaxed section constraint.
It is straightforward to verify that these additional terms in the transformation induce a modification
of the gauge invariant vector field strengths to 
\bea
{\cal F}_{\mu\nu}{}^M &\equiv& 2\,\partial_{[\mu} A_{\nu]}{}^M + 
10\,d^{MNK}\,\partial_N B_{\mu\nu\,K}
+ \sqrt{10}\,\Delta^{MK}\,\partial_\bullet B_{\mu\nu\,K}
\;,
\eea
as well an extension of the topological term, such that the combined vector-tensor couplings take the form
\bea
{\cal L}_{{\rm vt}, \bullet}  =  
-\frac{1}{4}\,\Delta_{MN}\,{\cal F}^{\mu\nu\,M}{\cal F}_{\mu\nu}{}^N
-\frac{5}{4}\,
\varepsilon^{\mu\nu\rho\sigma\tau}\,
  \partial_\mu B_{\nu\rho\,M}
\left(
\sqrt{10}\, d^{MNK}\partial_N B_{\sigma\tau\,K}
+\Delta^{MK} \partial_\bullet B_{\sigma\tau\,K}
\right)
\;,
\nonumber\\
\label{Lvtb}
\eea
and are invariant under these gauge transformations.
Let us work out the effect of these modifications for the exotic solutions of the
section constraint. With the kinetic scalar term unchanged, the resulting 
couplings are directly inferred from evaluating the covariant derivatives
(\ref{ExFTobjects}) for the $d$-symbol (\ref{dF4}), giving rise to
\bea
(3,1) &\!\longrightarrow\!&
 - \frac{1}{2}\left(\partial^\mu \phi^i -\partialsix  A^{\mu\,i} \right)
 \left(\partial_\mu \phi^i -\partialsix  A_\mu{}^{i} \right)
- \frac{1}{2}\,\partial^\mu \phi^\alpha \partial_\mu \phi^\alpha
\;,
\quad
i=1, \dots, 14\;,\;
\alpha=1, \dots, 28\;,
\nonumber\\
(4,0) &\!\longrightarrow\!&
 - \frac{1}{2}\,\partial^\mu \phi^A \partial_\mu \phi^A
 \;,\quad
A=1, \dots, 42
\;,
\eea
This precisely reproduces the vector-scalar couplings found in the explicit Lagrangians (\ref{L31full}), (\ref{L40full}) above.
As for the vector-tensor couplings, evaluating the Lagrangian (\ref{Lvtb}) with (\ref{dF4}) for the solutions (\ref{section_ext_sol})
gives rise to the explicit couplings
\bea
(3,1) &\!\!\longrightarrow\!\!&
-\frac{1}{4}\,(F_{\mu\nu} +\frac{3\sqrt{3}}{2} \, \partialsix  B_{\mu\nu} ) (F^{\mu\nu}  +\frac{3\sqrt{3}}{2} \, \partialsix  B^{\mu\nu} ) 
-\frac14\,\left( F_{\mu\nu}{}^a+ \partialsix  B_{\mu\nu}{}^a \right)\left( F^{\mu\nu\,a}+ \partialsix  B^{\mu\nu\,a} \right)
\nonumber\\
&&{}
 -\frac{1}{4}\, F^{\mu\nu\,i}F_{\mu\nu}{}^i 
-\frac{3}{16} \, \varepsilon^{\mu\nu\rho\sigma\tau} \partialsix  B_{\mu\nu} H_{\rho\sigma\tau} 
- \frac1{24}\,\varepsilon^{\mu\nu\rho\sigma\tau}\, \partialsix  B_{\mu\nu}{}^a \,H_{\rho\sigma\tau}{}^a
\;,
\nonumber\\[1ex]
(4,0) &\!\!\longrightarrow\!\!&
-\frac14\,\left( F_{\mu\nu}{}^M+ \partialsix  B_{\mu\nu}{}^M \right)\left( F^{\mu\nu\,M}+ \partialsix  B^{\mu\nu\,M} \right)
- \frac1{24}\,\varepsilon^{\mu\nu\rho\sigma\tau}\, \partialsix  B_{\mu\nu}{}^M \,H_{\rho\sigma\tau}{}^M
\;,
\eea
with indices in range $i=1, \dots, 14$, $a=1, \dots, 12$, $M=1, \dots, 27$, as above.
Again, this precisely reproduces the couplings found above (after proper rescaling of the vector field $A_\mu$)!

To summarize, in the scalar, vector and tensor sector, we have constructed an extension of the ExFT Lagrangian
(at the linearized level), given by
\bea
{\cal L} &=&
-\frac12\,D_\mu\phi^{MN} D^\mu \phi_{MN}
+{\cal L}_{{\rm vt}, \bullet}
+{\cal L}_{{\rm pot},\bullet}
\;,
\eea
which is invariant under the gauge transformations (\ref{gen_diff_lin}), (\ref{Lambda_mu})
modulo the relaxed section constraint (\ref{sectionext}). The weaker section constraint necessitates a numer of additional
contributions to the Lagrangian (and transformation rules) which precisely reproduce the explicit couplings 
found in the Lagrangians of the exotic models (\ref{L31full}), (\ref{L40full}) constructed above.
It is remarkable that this match confirms the couplings that have been determined from an underlying non-compact
E$_{6(6)}$ and F$_{4(4)}$ structure, respectively, despite the fact that the free theory only exhibits invariance under
the compact $R$-symmetry subgroup ${\rm USp}(2{\cal N}_+)\times {\rm USp}(2{\cal N}_-)$ which might in principle allow
for much more general couplings. We take this as evidence for the conjectured E$_{6(6)}$ and F$_{4(4)}$ invariance of the putative interacting theories~\cite{Hull:2000zn}.

\subsection{The spin-2 sector}
\label{subsec:spin2}

The above findings have revealed a very intriguing common structure of the couplings in the scalar, vector and tensor sectors of the 
different models which can be consistently embedded into an extension of (linearized) exceptional field theory. For the spin-2 sector 
carrying the Pauli-Fierz field and its duals on the other hand the picture appears not yet complete.
Extrapolation of the Lagrangian of the ${\cal N}=(4,0)$ model (\ref{L40}) suggests an extension of the standard ExFT Lagrangian
by couplings carrying $\partial_\bullet$ derivatives and the dual graviton fields as
 \bea
{\cal L}
&=&
-\frac14\,\widehat{\Omega}^{\mu\nu\rho}\widehat{\Omega}_{\mu\nu\rho}
+\frac12\,\widehat{\Omega}^{\mu\nu\rho} \widehat{\Omega}_{\nu\rho\mu}
+\widehat{\Omega}^{\mu}\widehat{\Omega}_{\mu}
+ \frac1{8}\,\varepsilon_{\mu\nu\sigma \kappa\lambda}\,
\partial^{\mu} \widehat{{C}}{}^{\nu\sigma}{}_{\rho} \,\partial_\bullet  \widehat{{C}}{}^{\kappa\lambda,\rho}
\nonumber\\
&&{}
- \frac1{32}\,\varepsilon_{\mu\nu\sigma \kappa\lambda}\,
\partial^{\mu} {\cal C}{}^{\nu\sigma}{}_{\rho} \,\partial_\bullet  {\cal C}{}^{\kappa\lambda,\rho}
+ \frac1{8}\,\partial_\bullet  {\cal C}{}_{\sigma\tau,\nu}\,\partial_{\mu} T^{\mu\nu,\sigma\tau}
-\frac1{4}\,\partial_\bullet  {\cal C}{}_{\kappa\lambda,\tau}\,\partial^{\kappa}  T^{\lambda\sigma,\tau}{}_{\sigma}
\nonumber\\
&&{}
- \frac1{4}\,\partial_{\nu} {\cal C}{}_{\sigma\mu}{}^{\mu}\,\partial_\bullet  T^{\sigma\tau,\nu}{}_{\tau}
+ \frac1{8}\,\partial_\bullet  {\cal C}{}_{\sigma\mu}{}^{\mu}\,\partial^{\sigma} T_{\tau\nu}{}^{\tau\nu}
+\frac1{64}\,\varepsilon_{\mu\nu\alpha\beta\gamma}\,
\partial^\alpha  T_{\sigma\tau}{}^{\beta\gamma}
\,\partial_\bullet  T^{\mu\nu,\sigma\tau}
\nonumber\\
&&{}
- \frac1{32}\,\partial_\bullet  T_{\sigma\tau,\mu\nu}\,\partial_\bullet  T^{\mu\nu,\sigma\tau}
+\frac18\,\partial_\bullet  T_{\sigma\mu,\nu}{}^\mu\,\partial_\bullet  T^{\sigma\tau,\nu}{}_{\tau}
- \frac1{32}\,\partial_\bullet  T_{\mu\nu}{}^{\mu\nu} \,\partial_\bullet  T_{\sigma\tau}{}^{\sigma\tau}
\nonumber\\
&&{}
+\frac{5}{4}\ \varepsilon^{\mu\nu\rho\sigma\tau}\,
d^{KMN} \partial_K {\cal B}_{\mu\nu\,M} \,\partial_N \widehat{C}_{\rho\sigma,\tau}
\;,
\label{spin2}
\eea
with
\bea
  \widehat{\Omega}_{\mu\nu\rho}
  &=&
  \partial_{[\mu} h_{\nu]\rho}  
  -\frac23\,\partial_M A_{[\mu}{}^M\,\eta_{\nu]\rho}
 -\partial_\bullet  \widehat{{C}}_{\mu\nu,\rho}
 +\frac{1}{2}\partial_\bullet  {\cal{C}}_{\mu\nu,\rho}
 \;.
\eea
By construction, this reproduces the ${\cal N}=(2,2)$ and the ${\cal N}=(4,0)$ models upon choosing the corresponding 
solutions of the section constraint. It remains unclear however, how the spin-2 sector of the ${\cal N}=(3,1)$ model can find its place
in this construction. In particular, the appearance of the extra fields ${\cal C}_{\mu\nu,\rho}$ and $T_{\mu\nu,\rho\sigma}$
appearing in (\ref{spin2}), whose couplings remain present upon selecting the (3,1) solution (\ref{section_ext_sol})
of the section constraint, poses a challenge for recovering the Lagrangian (\ref{L31full}) of the ${\cal N}=(3,1)$ model. 
The structure of the gauge transformations of ${\cal C}$ as extrapolated from (\ref{gauge40OCC}) appears to suggest
a gauge fixing of the $\zeta_{\mu\nu}$ and $\lambda_{\rho,\mu\nu}$ gauge symmetries --- absent in the ${\cal N}=(3,1)$ model ---
in order to remove this field. Another apparent problem in the spin-2 sector is the lacking reconciliation 
between the $\partial_\bullet h\partial_\bullet h$ 
terms from (\ref{pot_ext}) and the $\partial_\bullet T\partial_\bullet T$ terms of (\ref{spin2}) which mutually violate
the correct limits to the exotic models.
Resolution of this problem may require to implement algebraic relations between the 
Pauli-Fierz $h_{\mu\nu}$ field and the double dual graviton \cite{Hull:2000rr} (see also \cite{Henneaux:2019zod}).

\section{Conclusions and Outlook} 

In this paper we have taken the first step in constructing action principles for exotic supergravity theories in 6D 
by giving such actions for the free bosonic part. These actions show already intriguing new features such as 
the simultaneous appearance of (linearized) diffeomorphisms and dual diffeomorphisms, which are realized
on exotic Young tableaux fields as well as more conventional gravity fields. Our formulation abandons manifest 6D 
Lorentz invariance, as expected to be necessary on general grounds, by being based on a $5+1$ split of coordinates. 
Remarkably, the field equations implied by our actions can be integrated to reconstruct  the 
correct dynamics of these exotic supergravites. 
Moreover, we have seen the first glimpses of an exceptional field theory master formulation, 
in which the conventional ${\cal N}=(2,2)$, as well as the exotic ${\cal N}=(3,1)$ and ${\cal N}=(4,0)$ models 
all emerge through different solutions of an extended section constraint, 
but clearly much more needs to be done. We close with a brief discussion of possible future developments. 

First, it remains to exhibit the (maximal) supersymmetries in these non-standard formulations, even just at the free level. 
We have no doubt that this can be achieved as in exceptional field theory where different 
supersymmetries (such as type IIB versus type IIA) are realized within a single master formulation. Second, it would 
be interesting to study possible  embeddings into exceptional field theories of higher rank, such as for U-duality groups E$_{7(7)}$ 
and E$_{8(8)}$, which may illuminate some issues and which can also be done already at linearized level. 
Finally, the most important outstanding problem is clearly the question whether our formulation can be extended to 
the non-linear interacting  theory. We would like to emphasize that the present formulations seem quite promising in this regard 
since they feature not only the exotic fields but also the more conventional gravity fields, which come with an action that
allows a natural embedding into the full non-linear Einstein-Hilbert action.
In turn this suggests 
that all these fields might become part of a  tensor hierarchy that extends to the gravity sector. If so this could quite naturally 
lend itself to a formulation of non-linear dynamics in terms of a hierarchy of duality relations as in \cite{Bonezzi:2019bek}.

\paragraph{Acknowledgements}
We wish to thank X. Bekaert, C. Hull and V. Lekeu for enlightening discussions.
The work of O.H. is supported by the ERC Consolidator Grant ``Symmetries \& Cosmology''.

\begin{appendix}

\section*{Appendix}

\section{Actions for selfdual tensor fields}
\label{app:HTExFT}

It is well known that the first-order field equations for $D=6$ selfdual tensor fields
\bea
H_{\hat\mu\hat\nu\hat\rho} &=& 
\frac16\,\varepsilon_{\hat\mu\hat\nu\hat\rho\hat\sigma\hat\kappa\hat\lambda}\,H^{\hat\sigma\hat\kappa\hat\lambda}
\;,
\qquad
H_{\hat\mu\hat\nu\hat\rho}  = 3\,\partial_{[\hat\mu} B_{\hat\nu\hat\rho]}
\;,
\label{6DsdAPP}
\eea
do not integrate to a standard action principle, yet various mechanisms with different characteristics have been devised such as to provide
a Lagrangian description of these equations~\cite{Henneaux:1988gg,Pasti:1996vs,Sen:2019qit,Mkrtchyan:2019opf}.
In this appendix, we briefly review the construction of Henneaux and Teitelboim \cite{Henneaux:1988gg} which is somewhat closest in spirit
to the construction employed in this paper, together with its dual formulation that is naturally embedded within exceptional field theory.
Both formulations are based on a coordinate split (\ref{split51})
\bea
\{ x^{\hat{\mu}}\} \longrightarrow \{x^\mu, y \}
\;,
\label{split51APP}
\eea 
and sacrifice manifest $D=6$ Poincar\'e invariance.\footnote{
The original construction of \cite{Henneaux:1988gg} defines the $5+1$ split
by singling out the time coordinate, but the method obviously applies equally well for the split based on a spatial distinguished dimension.
}
With the corresponding split $\{B_{\hat\mu\hat\nu}\}=\{B_{\mu\nu}, B_{\mu 5} \equiv A_\mu\}$ of the six-dimensional tensor field,
the selfduality equations (\ref{6DsdAPP}) take the form
\bea
{\cal F}_{\mu\nu} + \frac16 \varepsilon_{\mu\nu\rho\sigma\tau}\,H^{\rho\sigma\tau} &=& 0
\;,\qquad
\mbox{for}\quad
{\cal F}_{\mu\nu} \equiv  {F}_{\mu\nu}+\partialsix B_{\mu\nu}
\;.
\label{sd}
\eea
In particular, the divergence and curl of this equation give rise to
\bea
\partial^\mu {\cal F}_{\mu\nu}  &=& 0 \;,
\nonumber\\
\varepsilon^{\mu\nu\lambda \sigma\tau} \partialsix H_{\lambda\sigma\tau} 
- 6\,\partial_\lambda H^{\lambda \mu\nu} &=& 0
\;,
\label{Bianchisd}
\eea
respectively.

\subsection{Henneaux-Teitelboim Lagrangian}
\label{app:HT}

The Lagrangian proposed by Henneaux and Teitelboim~\cite{Henneaux:1988gg} for the description of the self-dual tensors
takes the form
\bea
{\cal L} &=& \frac1{24}\,\varepsilon^{\mu\nu\rho\sigma\tau}\,{\cal F}_{\mu\nu}\,H_{\rho\sigma\tau}
-\frac1{12}\,H_{\mu\nu\rho}\,H^{\mu\nu\rho}
\;,
\label{HTsd}
\eea
when applied to equations (\ref{sd}), i.e.\ evaluated for space-like split and flat background. As a first observation, 
this Lagrangian depends on the vector field $A_\mu$ only via total derivatives, such that it does not
show up in the field equations
\bea
\varepsilon_{\mu\nu\rho\sigma\tau}\, \partialsix H^{\rho\sigma\tau} &=&
6\,\partial^\rho H_{\mu\nu\rho}
\;,
\label{eomHT}
\eea
reproducing the second equation of (\ref{Bianchisd}). 
This equation now serves an an integrability equation in order to locally define the vector field $A_\mu$
via the equation
\bea
2\,\partial_{[\mu} A_{\nu]} &=& 
-\partialsix B_{\mu\nu} -  \frac16 \varepsilon_{\mu\nu\rho\sigma\tau}\,H^{\rho\sigma\tau} 
\;.
\label{defAeq}
\eea
Indeed, the curl of the r.h.s.\ vanishes by virtue of (\ref{eomHT}).
Defining the vector field $A_\mu$ by (\ref{defAeq}), we precisely recover the equations of motion (\ref{sd}).

\subsection{ExFT type Lagrangian}
\label{app:ExFT}

Exceptional field theory (ExFT) typically yields formulations of higher-dimensional supergravity 
theories based on the field content of lower-dimensional theories. In particular, it offers actions for 
theories that do not admit actions in terms of their original variables, such as IIB supergravity, c.f.~\cite{Baguet:2015xha}.
In the context of (anti-)selfdual tensor fields appearing in six dimensions, an exceptional field theory formulation
based on a split (\ref{split51APP}) gives rise to an action
\bea
{\cal L} &=& 
-\frac14\,{\cal F}_{\mu\nu}{\cal F}^{\mu\nu}
-\frac1{24}\,\varepsilon^{\mu\nu\rho\sigma\tau}\,\partialsix B_{\mu\nu}\,H_{\rho\sigma\tau}
\;,
\label{ExFTsd}
\eea
carrying the fields of equation (\ref{sd}). 
The field equations are now given by
\bea
0 &=&
\partial^\nu  {\cal F}_{\nu\mu}
~=~
\partial^\nu  {F}_{\nu\mu} + \partialsix \partial^\nu  B_{\nu\mu} \;,
\label{eomExFTds1}\\
0 &=& 
\partialsix \Big(
{\cal F}_{\mu\nu}
+
\frac16\,\varepsilon_{\mu\nu\rho\sigma\tau}\, H^{\rho\sigma\tau} \Big)
\;.
\label{eomExFTds2}
\eea
In particular, equation (\ref{eomExFTds2}) implies the original field equations (\ref{sd}) up to some
function that does not depend on $y$:
\bea
{\cal F}_{\mu\nu}
+
\frac16\,\varepsilon_{\mu\nu\rho\sigma\tau}\, H^{\rho\sigma\tau}
&=& \chi_{\mu\nu}\;,\qquad
\partialsix\chi_{\mu\nu}=0
\;.
\label{eomExFT3}
\eea
Comparing the divergence of this equation to (\ref{eomExFTds1}), we find that locally the field $\chi_{\mu\nu}$ can be integrated to
\bea
\partial^\mu \chi_{\mu\nu} &=& 0
\quad
\Longrightarrow\quad
\chi_{\mu\nu} = \varepsilon_{\mu\nu\rho\sigma\tau} \,\partial^\rho b^{\sigma\tau}
\;,
\eea
in terms of a function $b_{\mu\nu}$, such that the field equations (\ref{eomExFT3}) can be rewritten as
\bea
\left(
{F}_{\mu\nu}+\partialsix \tilde{B}_{\mu\nu}
\right)
+
\frac16\,\varepsilon_{\mu\nu\rho\sigma\tau}\, 
3\, \partial^\rho \tilde B^{\sigma\tau} 
&=& 0
\;,
\label{eomExFT4}
\eea
with the modified two-form
\bea
\tilde{B}_{\mu\nu} &\equiv& B_{\mu\nu} -2\,b_{\mu\nu}
\;.
\label{modB}
\eea
In terms of the fields $A_\mu$, $\tilde{B}_{\mu\nu}$, we thus recover the
desired original field equations (\ref{sd}).
Note finally, that the Lagrangian (\ref{ExFTsd}) precisely comes with a gauge freedom of the type (\ref{modB})
which allows to absorb $b_{\mu\nu}$ into $B_{\mu\nu}$\,.

We thus arrive at two complementary Lagrangians (\ref{HTsd}), (\ref{ExFTsd}), which both describe the six-dimensional selfdual
tensor field upon sacrificing manifest $D=6$ Poincar\'e invariance. They are dual to each other in the sense that upon dimensional
reduction to $D=5$ dimensions, i.e.\ upon setting $\partialsix\rightarrow0$, the Lagrangian (\ref{ExFTsd}) describes the 3 degrees
of freedom in terms of a free Maxwell field whereas (\ref{HTsd}) describes them in terms of the dual massless tensor field $B_{\mu\nu}$\,.
Similarly, the two Lagrangians (\ref{HTsd}) and (\ref{ExFTsd}) can be dualized into each other in presence of the sixth dimension.

\section{6D field equations from the new Lagrangians}
\label{app:31}

In this appendix, we present in detail how the second-order field equations obtained by variation of the Lagrangian 
(\ref{L31}) for the ${\cal N}=(3,1)$ model can be integrated to the first-order field equations (\ref{dualFH6}) and (\ref{defc2})
which in turn imply the original 6D second-order selfduality equations (\ref{6DsdC31}).
For the ${\cal N}=(4,0)$ model (\ref{L40}), the discussion goes along the same lines.

\subsection{Field equations}
\label{subsec:eom31}

Here, we spell out the field equations obtained from variation of the Lagrangian (\ref{L31}).

Variation w.r.t. $A_\mu$:
\begin{equation}
\partial^\mu F_{\mu\nu} + \frac{3}{2}\, \partialsix \partial^\mu B_{\mu\nu} -\frac12\, \partialsix \left( \partial^\mu h_{\mu\nu} - \partial_\nu h_\mu{}^\mu - \partialsix \widehat{C}_{\nu\mu}{}^\mu + 4\, \partialsix A_\nu \right) =0,
\label{varA}
\end{equation}
which is exactly (\ref{Maxwell31}).

Variation w.r.t. $B_{\mu\nu}$:
\begin{equation}
\partialsix \left( F_{\mu\nu} + \frac{3}{2}\, \partialsix B_{\mu\nu} + \frac{1}{2}\, \varepsilon_{\mu\nu\rho\sigma\tau} \partial^\rho B^{\sigma\tau} + \frac{1}{12}\, \partialsix \varepsilon_{\mu\nu\rho\sigma\tau} \widehat{C}^{\rho\sigma\tau} \right) = 0,
\label{varB}
\end{equation}
which is the $\partialsix$ derivative of equation (\ref{dualFH6}).

Variation w.r.t. $h_{\mu\nu}$:
\begin{equation}
{\cal G}_{\mu\nu}+ \frac{1}{2}\, \partialsix \left( \partial^\rho \widehat{C}_{\rho (\mu,\nu)} + \partial_{(\mu} \widehat{C}_{\nu)\rho}{}^\rho - \eta_{\mu\nu} \partial^\rho \widehat{C}_{\rho\sigma}{}^\sigma \right) =0,
\label{varh}
\end{equation}
with the linearized Einstein tensor as it appears in (\ref{Einstein31}), this variation thus exactly reproduces the Einstein equation (\ref{Einstein31}).

Variation w.r.t $\widehat{C}_{\mu\nu,\rho}$:
\begin{align}
\partialsix &\left( \partial_{[ \mu} h_{\nu]\rho} + \partial^\sigma h_{\sigma[ \mu} \eta_{\nu]\rho} - \eta_{\rho[ \nu}\partial_{\mu]} h_\sigma{}^\sigma  + \frac{1}{4}\,\varepsilon_{\mu\nu\lambda\sigma\tau} \partial^\lambda \widehat{C}^{\sigma\tau}{}_\rho + \frac{1}{4}\, \partialsix ( \widehat{C}_{\mu\nu,\rho} -\widehat{C}_{\nu\rho,\mu} - \widehat{C}_{\rho\mu,\nu} )  \right. \nonumber \\
&\qquad \left.   - \eta_{\rho[ \nu} \partialsix \widehat{C}_{\mu]\sigma}{}^\sigma + 3\, \partialsix A_{[ \mu} \eta_{\nu]\rho} + \frac{3}{8}\, \varepsilon_{\mu\nu\rho\sigma\tau} B^{\sigma\tau} \right)=0,
\label{varC}
\end{align}
which we can further project onto its (2,1) part and totally antisymmetric part
\begin{align}
\partialsix &\left( \partial_{[ \mu} h_{\nu]\rho} + \partial^\sigma h_{\sigma[ \mu} \eta_{\nu]\rho} - \eta_{\rho[ \nu}\partial_{\mu]} h_\sigma{}^\sigma  + \frac{1}{4}\,\varepsilon_{\mu\nu\lambda\sigma\tau} \partial^\lambda \widehat{C}^{\sigma\tau}{}_\rho - \frac{1}{4}\,\varepsilon_{\lambda\sigma\tau[\mu\nu} \partial^\lambda \widehat{C}^{\sigma\tau}{}_{\rho]}   \right. \nonumber \\
&\qquad \left.  + \frac{1}{4}\, \partialsix ( \widehat{C}_{\mu\nu,\rho} -\widehat{C}_{\nu\rho,\mu} - \widehat{C}_{\rho\mu,\nu} +  \widehat{C}_{[\mu\nu,\rho]})  - \eta_{\rho[ \nu} \partialsix \widehat{C}_{\mu]\sigma}{}^\sigma + 3\, \partialsix A_{[ \mu} \eta_{\nu]\rho} \right) =0, \\[1ex]
\partialsix & \left(  \frac{1}{4}\,\varepsilon_{\lambda\sigma\tau[\mu\nu} \partial^\lambda \widehat{C}^{\sigma\tau}{}_{\rho]} - \frac{1}{4}\, \partialsix \widehat{C}_{[\mu\nu,\rho]} + \frac{3}{8}\, \partialsix \varepsilon_{\mu\nu\rho\sigma\tau} B^{\sigma\tau} \right) =0.
\end{align}

\subsection{Going back to the original equations}

The goal of this section is to recover the full 6D system (\ref{defc}) and (\ref{dualFH6}) from the equations derived in the section~\ref{subsec:eom31}.
Let us first rewrite them in terms of the original fields of the (3,1) model and integrate all the equations under $\partialsix$ by introducing three functions $\chi_{\mu\nu} (x^\mu)$, $ \psi_{\mu\nu\rho}(x^\mu)$ and $\varphi_{\mu\nu,\rho}(x^\mu)$ which are respectively antisymmetric, antisymmetric and of $(2,1)$ type,
and do not depend on the sixth coordinate.
\begin{align}
&\partial^\mu F_{\mu\nu} + \frac{3}{2}\, \partialsix \partial^\mu B_{\mu\nu} -\frac12\, \partialsix \left( \partial^\mu h_{\mu\nu} - \partial_\nu h_\mu{}^\mu - \partialsix C_{\nu\mu}{}^\mu + 4\, \partialsix A_\nu \right) =0, \label{intvar1}\\
&F_{\mu\nu} + \frac{3}{2}\, \partialsix B_{\mu\nu} - \partialsix u_{\mu\nu} + \frac{1}{2}\, \varepsilon_{\mu\nu\rho\sigma\tau} \partial^\rho B^{\sigma\tau} = \chi_{\mu\nu}, \label{intvar2}\\
& \varepsilon_{\lambda\sigma\tau[\mu\nu} \partial^\lambda C^{\sigma\tau}{}_{\rho]} - 4\,\partial_{[\rho}u_{\mu\nu]} - \partialsix \varepsilon_{\mu\nu\rho\sigma\tau}  \left( u^{\sigma\tau} -\frac{3}{2}\,B^{\sigma\tau}\right) = \psi_{\mu\nu\rho}, \label{intvar3}\\
& \mathcal{G}_{\mu\nu} + \frac{1}{2}\partialsix \left( \partial^\rho C_{\rho (\mu,\nu)} + \partial_{(\mu} C_{\nu)\rho}{}^\rho - \eta_{\mu\nu} \partial^\rho C_{\rho\sigma}{}^\sigma  \right)=0, \label{intvar4}\\
& \partial_{[ \mu} h_{\nu]\rho} + \partial^\sigma h_{\sigma[ \mu} \eta_{\nu]\rho} - \eta_{\rho[ \nu}\partial_{\mu]} h_\sigma{}^\sigma  + \frac{1}{4}\,\varepsilon_{\mu\nu\lambda\sigma\tau} \partial^\lambda C^{\sigma\tau}{}_\rho -  \frac{1}{4}\,\varepsilon_{\lambda\sigma\tau[ \mu\nu} \partial^\lambda C^{\sigma\tau}{}_{\rho]} \nonumber \\
 &\quad + \frac{1}{2}\, \partialsix C_{\mu\nu,\rho} - \eta_{\rho[ \nu} \partialsix C_{\mu]\sigma}{}^\sigma + 3\, \partialsix A_{[ \mu} \eta_{\nu]\rho} - \partial_\rho u_{\mu\nu}  + \partial_{[ \rho} u_{\mu\nu]} - 2\, \partial^\sigma u_{\sigma [ \mu} \eta_{\nu]\rho} =\varphi_{\mu\nu,\rho}.\label{intvar5}
\end{align}

\paragraph{A-B duality}
Combining (\ref{intvar1}) and (\ref{intvar2}) gives
\begin{equation}
\partialsix\partial^\mu u_{\mu\nu} = \frac12\, \partialsix \left( \partial^\mu h_{\mu\nu} - \partial_\nu h_\mu{}^\mu - \partialsix C_{\nu\mu}{}^\mu + 4\, \partialsix A_\nu  \right) - \partial^\mu \chi_{\mu\nu},
\end{equation}
while the trace of (\ref{intvar5}) in $(\mu\rho)$ gives
\begin{equation}
\partial^\mu u_{\mu\nu} = \frac12\, \left( \partial^\mu h_{\mu\nu} - \partial_\nu h_\mu{}^\mu - \partialsix C_{\nu\mu}{}^\mu + 4\, \partialsix A_\nu  \right) + \frac{1}{3}\,\varphi_{\mu\nu}{}^\mu. \label{intvar5trace}
\end{equation}
Together, these two equations imply that locally, we can define a 2-form $b$ such that
\begin{equation}
\chi_{\mu\nu} = \frac{1}{2}\,\varepsilon_{\mu\nu\rho\sigma\tau}\partial^\rho b^{\sigma\tau} (x^\mu).
\end{equation}
This 2-form can be absorbed in $B$ (following exactly the same process as in section \ref{app:ExFT}) 
such that equations (\ref{intvar2}) reproduces (\ref{dualFH6}).

\paragraph{h-C duality}

Contracting (\ref{intvar5}) with $\partial^\mu$, we can extract both symmetric and antisymmetric parts:
\be
(\nu\rho):&\quad {\cal G}_{\nu\rho} + \frac{1}{2}\, \partialsix \left( \partial^\mu C_{\mu (\nu,\rho)} - \eta_{\nu\rho} \partial^\mu C_{\mu\sigma}{}^\sigma + \partial_{(\rho}C_{\nu)\sigma}{}^\sigma \right) =  \partial^\mu \varphi_{\mu (\nu,\rho)}, \label{intvar5divsym}\\
[\nu\rho]:&\quad  -\frac{1}{6}\, \varepsilon_{\lambda\sigma\tau\nu\rho} \partial^\mu \partial^\lambda C^{\sigma\tau}{}_\mu + \partialsix \left(\partial^\mu C_{\mu [\nu,\rho]} + \partial_{[\rho}C_{\nu]\sigma}{}^\sigma - 3\, \partial_{[\rho} A_{\nu]}\right) + 2\,\partial^\mu \partial_{[\mu}u_{\nu\rho]} = 2\,\partial^\mu \varphi_{\mu [\nu,\rho]}.  \label{intvar5divantisym}
\ee
Using (\ref{intvar4}) we can conclude that 
\begin{equation}
\partial^\mu \varphi_{\mu (\nu,\rho)}=0. \label{condphi}
\end{equation}
The divergence of (\ref{intvar3}) reads
\begin{equation}
-\frac{1}{6}\, \varepsilon_{\mu\nu\alpha\beta\gamma} \partial^\rho \partial^\alpha C^{\alpha\beta}{}_{\rho} + 2\, \partial^\rho \partial_{[\rho}u_{\mu\nu]} + \frac{1}{2}\, \partialsix \varepsilon_{\mu\nu\rho\alpha\beta}\partial^\rho \left(u^{\alpha\beta} - \frac{3}{2}\, B^{\alpha\beta}\right) = -\frac{1}{2}\, \partial^\rho \psi_{\mu\nu\rho},
\end{equation}
and combining it with (\ref{intvar5divantisym}) and (\ref{intvar2}), we eventually get
\begin{equation}
2\, \partial^\mu \varphi_{\mu[\nu,\rho]} = -\frac{1}{2}\, \partial^\rho \psi_{\mu\nu\rho}.
\end{equation}
Together with (\ref{condphi}), one has
\begin{equation}
\partial^\mu (2\, \varphi_{\mu\nu,\rho} +\frac{1}{2}\, \partial^\rho \psi_{\mu\nu\rho}) =0,
\end{equation}
such that locally there exist 5D tensors 
$c_{\mu\nu,\rho}$ and $a_{\mu\nu\rho}$, where $c$ is of $(2,1)$ type and $a$ is completely antisymmetric, such that
\begin{equation}
2\,\varphi_{\mu\nu,\rho} + \frac{1}{2}\, \partial^\rho \psi_{\mu\nu\rho} = \frac{1}{2}\, \varepsilon_{\mu\nu\alpha\beta\gamma}\partial^\alpha (c^{\beta\gamma}{}_\rho + a^{\beta\gamma}{}_\rho ).
\end{equation}
Consequently
\begin{align}
\varphi_{\mu\nu}{}^\mu &= \frac{1}{4}\,  \varepsilon_{\mu\nu\alpha\beta\gamma}\partial^\alpha a^{\beta\gamma\mu}\;, \\
\varphi_{\mu\nu,\rho} &= \frac{1}{4}\, \varepsilon_{\mu\nu\alpha\beta\gamma}\partial^\alpha \Big(c^{\beta\gamma}{}_\rho +  a^{\beta\gamma}{}_\rho \Big) - \frac{1}{4}\, \varepsilon_{\alpha\beta\gamma[\mu\nu}\partial^\alpha \Big(c^{\beta\gamma}{}_{\rho]} + a^{\beta\gamma}{}_{\rho]} \Big)\;, \\
\psi_{\mu\nu\rho} &= \varepsilon_{\alpha\beta\gamma[\mu\nu}\partial^\alpha \Big(c^{\beta\gamma}{}_{\rho]} + a^{\beta\gamma}{}_{\rho]} \Big)\;.
\end{align}
Plugging the expression for $\varphi$ and its trace back into (\ref{intvar5}), one has
\begin{align}
2\, \partial_{[ \mu} h_{\nu]\rho} &+ \frac{1}{2}\,\varepsilon_{\mu\nu\lambda\sigma\tau} \partial^\lambda C^{\sigma\tau}{}_\rho -  \frac{1}{2}\,\varepsilon_{\lambda\sigma\tau[ \mu\nu} \partial^\lambda C^{\sigma\tau}{}_{\rho]} + \partialsix C_{\mu\nu,\rho} - 2\, \partialsix A_{[ \mu} \eta_{\nu]\rho} - 2\,\partial_\rho u_{\mu\nu}  + 2\,\partial_{[ \rho} u_{\mu\nu]}\nonumber \\
&=\frac{1}{2}\, \varepsilon_{\mu\nu\alpha\beta\gamma}\partial^\alpha (c^{\beta\gamma}{}_\rho +  a^{\beta\gamma}{}_\rho ) - \frac{1}{2}\, \varepsilon_{\alpha\beta\gamma[\mu\nu}\partial^\alpha (c^{\beta\gamma}{}_{\rho]} + a^{\beta\gamma}{}_{\rho]} ) + \frac{1}{3}\,\varepsilon_{\alpha\beta\gamma\sigma[\mu}\partial^\alpha a^{\beta\gamma\sigma} \eta_{\nu]\rho}\;.
\label{varC2}
\end{align}
Then, using the following two Schouten identities
\begin{align}
\varepsilon_{[\mu\nu}{}^{\alpha\beta\gamma}\partial_\alpha a_{\beta\gamma\rho]} &= 0 = - \varepsilon^{\alpha\beta\gamma}{}_{[\mu\nu}\partial_{\rho]} a_{\alpha\beta\gamma} + 3\, \partial_\alpha \varepsilon^{\alpha\beta\gamma}{}_{[\mu\nu} a_{\rho]\beta\gamma}\;, \\
\varepsilon_{[\sigma\mu\alpha\beta\gamma}\partial^\alpha a^{\beta\gamma\sigma}\eta_{\nu]\rho} &= 0 = 2\,\varepsilon_{\alpha\beta\gamma\sigma[\mu}\partial^\alpha a^{\beta\gamma\sigma}\eta_{\nu]\rho} - \partial_\rho \varepsilon_{\beta\gamma\sigma\mu\nu} a^{\beta\gamma\sigma} + 3\,  \varepsilon_{\alpha\beta\gamma\mu\nu}\partial^\alpha a^{\beta\gamma}{}_\rho\;,
\end{align}
one obtains
\begin{align}
2\, \partial_{[ \mu} h_{\nu]\rho} + \frac{1}{2}\,\varepsilon_{\mu\nu\lambda\sigma\tau} \partial^\lambda (C^{\sigma\tau}{}_\rho - c^{\sigma\tau}{}_\rho) +( H_{\mu\nu\rho} - \frac{1}{2}\, \varepsilon_{\mu\nu\rho\alpha\beta} F^{\alpha\beta}) + \partialsix (C_{\mu\nu,\rho} - 2\,  A_{[ \mu} \eta_{\nu]\rho})& \nonumber \\
 - 2\,\partial_\rho \left(u_{\mu\nu} + \frac{1}{12}\, \varepsilon_{\mu\nu\alpha\beta\gamma}a^{\alpha\beta\gamma}\right) &= 0\;.
\end{align}
We recover the 6D equation (\ref{defc}) after the following redefinitions
\begin{align}
u_{\mu\nu} &\to u_{\mu\nu} + \frac{1}{12}\, \varepsilon_{\mu\nu\alpha\beta\gamma}a^{\alpha\beta\gamma}\;,\nonumber \\
B_{\mu\nu} &\to B_{\mu\nu} - b_{\mu\nu}\;, \quad
C_{\mu\nu,\rho} \to C_{\mu\nu,\rho}  - c_{\mu\nu,\rho}\;.
\end{align}
One can check that these redefinitions are consistent with the expression of $\psi$ in (\ref{intvar3}).
Finally, derivative of (\ref{defc}) and (\ref{dualFH6}) give rise to the original 6D equations of motion (\ref{6DsdC31})
as discussed in section~\ref{subsec:split31} above.

\end{appendix}



\begin{thebibliography}{10}

\bibitem{Hull:2000zn}
C.~Hull, ``Strongly coupled gravity and duality,''
  \href{http://dx.doi.org/10.1016/S0550-3213(00)00323-0}{{\em Nucl.Phys.}
  {\bfseries B583} (2000) 237--259},
\href{http://arxiv.org/abs/hep-th/0004195}{{\ttfamily arXiv:hep-th/0004195
  [hep-th]}}.

\bibitem{Hull:2000rr}
C.~M. Hull, ``Symmetries and compactifications of $(4,0)$ conformal gravity,''
  \href{http://dx.doi.org/10.1088/1126-6708/2000/12/007}{{\em JHEP} {\bfseries
  12} (2000) 007},
\href{http://arxiv.org/abs/hep-th/0011215}{{\ttfamily arXiv:hep-th/0011215
  [hep-th]}}.

\bibitem{Chiodaroli:2011pp}
M.~Chiodaroli, M.~G\"unaydin, and R.~Roiban, ``Superconformal symmetry and
  maximal supergravity in various dimensions,''
  \href{http://dx.doi.org/10.1007/JHEP03(2012)093}{{\em JHEP} {\bfseries 03}
  (2012) 093}, \href{http://arxiv.org/abs/1108.3085}{{\ttfamily arXiv:1108.3085
  [hep-th]}}.

\bibitem{Anastasiou:2013hba}
A.~Anastasiou, L.~Borsten, M.~Duff, L.~Hughes, and S.~Nagy, ``A magic pyramid
  of supergravities,'' \href{http://dx.doi.org/10.1007/JHEP04(2014)178}{{\em
  JHEP} {\bfseries 04} (2014) 178},
  \href{http://arxiv.org/abs/1312.6523}{{\ttfamily arXiv:1312.6523 [hep-th]}}.

\bibitem{Borsten:2017jpt}
L.~Borsten, ``{$D=6$}, {$\mathcal{N}=(2,0)$} and {$\mathcal{N}=(4,0)$}
  theories,'' \href{http://dx.doi.org/10.1103/PhysRevD.97.066014}{{\em Phys.
  Rev. D} {\bfseries 97} no.~6, (2018) 066014},
  \href{http://arxiv.org/abs/1708.02573}{{\ttfamily arXiv:1708.02573
  [hep-th]}}.

\bibitem{Strathdee:1986jr}
J.~A. Strathdee, ``Extended {P}oincar\'e supersymmetry,''
\href{http://dx.doi.org/10.1142/S0217751X87000120}{{\em Int. J. Mod. Phys.}
  {\bfseries A2} (1987) 273}.

\bibitem{Hull:2000cf}
C.~Hull, ``{BPS} supermultiplets in five-dimensions,''
  \href{http://dx.doi.org/10.1088/1126-6708/2000/06/019}{{\em JHEP} {\bfseries
  06} (2000) 019}, \href{http://arxiv.org/abs/hep-th/0004086}{{\ttfamily
  arXiv:hep-th/0004086}}.

\bibitem{Berman:2010is}
D.~S. Berman and M.~J. Perry, ``Generalized geometry and {M} theory,''
  \href{http://dx.doi.org/10.1007/JHEP06(2011)074}{{\em JHEP} {\bfseries 1106}
  (2011) 074},
\href{http://arxiv.org/abs/1008.1763}{{\ttfamily arXiv:1008.1763 [hep-th]}}.

\bibitem{Coimbra:2011ky}
A.~Coimbra, C.~Strickland-Constable, and D.~Waldram, ``{$E_{d(d)} \times
  \mathbb{R}^+$ generalised geometry, connections and M theory},''
  \href{http://dx.doi.org/10.1007/JHEP02(2014)054}{{\em JHEP} {\bfseries 1402}
  (2014) 054},
\href{http://arxiv.org/abs/1112.3989}{{\ttfamily arXiv:1112.3989 [hep-th]}}.

\bibitem{Hohm:2013pua}
O.~Hohm and H.~Samtleben, ``Exceptional form of ${D}=11$ supergravity,''
  \href{http://dx.doi.org/10.1103/PhysRevLett.111.231601}{{\em Phys. Rev.
  Lett.} {\bfseries 111} (2013) 231601},
\href{http://arxiv.org/abs/1308.1673}{{\ttfamily arXiv:1308.1673 [hep-th]}}.

\bibitem{Hohm:2013vpa}
O.~Hohm and H.~Samtleben, ``Exceptional field theory {I}: {E}$_{6(6)}$
  covariant form of {M}-theory and type {IIB},''
  \href{http://dx.doi.org/10.1103/PhysRevD.89.066016}{{\em Phys.Rev.}
  {\bfseries D89} (2014) 066016},
\href{http://arxiv.org/abs/1312.0614}{{\ttfamily arXiv:1312.0614 [hep-th]}}.

\bibitem{Baguet:2015xha}
A.~Baguet, O.~Hohm, and H.~Samtleben, ``{E$_{6(6)}$} exceptional field theory:
  Review and embedding of type {IIB},'' in {\em {14th Hellenic School and
  Workshops on Elementary Particle Physics and Gravity Corfu, 2014}}, vol.~PoS
  (CORFU2014), p.~133.
\newblock 2015.
\newblock
\href{http://arxiv.org/abs/1506.01065}{{\ttfamily arXiv:1506.01065 [hep-th]}}.
\newblock

\bibitem{Henneaux:1988gg}
M.~Henneaux and C.~Teitelboim, ``Dynamics of chiral (self-dual) $p$-forms,''
\href{http://dx.doi.org/10.1016/0370-2693(88)90712-5}{{\em Phys. Lett.}
  {\bfseries B206} (1988) 650}.

\bibitem{Henneaux:2016opm}
M.~Henneaux, V.~Lekeu, and A.~Leonard, ``Chiral tensors of mixed {Y}oung
  symmetry,'' \href{http://dx.doi.org/10.1103/PhysRevD.95.084040}{{\em Phys.
  Rev. D} {\bfseries 95} no.~8, (2017) 084040},
  \href{http://arxiv.org/abs/1612.02772}{{\ttfamily arXiv:1612.02772
  [hep-th]}}.

\bibitem{Henneaux:2017xsb}
M.~Henneaux, V.~Lekeu, and A.~Leonard, ``The action of the (free) (4,
  0)-theory,'' \href{http://dx.doi.org/10.1007/JHEP01(2018)114}{{\em JHEP}
  {\bfseries 01} (2018) 114}, \href{http://arxiv.org/abs/1711.07448}{{\ttfamily
  arXiv:1711.07448 [hep-th]}}. [Erratum: JHEP 05, 105 (2018)].

\bibitem{Henneaux:2018rub}
M.~Henneaux, V.~Lekeu, J.~Matulich, and S.~Prohazka, ``The action of the (free)
  {$\mathcal{N} = (3,1)$} theory in six spacetime dimensions,''
  \href{http://dx.doi.org/10.1007/JHEP06(2018)057}{{\em JHEP} {\bfseries 06}
  (2018) 057}, \href{http://arxiv.org/abs/1804.10125}{{\ttfamily
  arXiv:1804.10125 [hep-th]}}.

\bibitem{Henneaux:2004jw}
M.~Henneaux and C.~Teitelboim, ``Duality in linearized gravity,''
  \href{http://dx.doi.org/10.1103/PhysRevD.71.024018}{{\em Phys. Rev. D}
  {\bfseries 71} (2005) 024018},
  \href{http://arxiv.org/abs/gr-qc/0408101}{{\ttfamily arXiv:gr-qc/0408101}}.

\bibitem{Minasian:2020vxn}
R.~Minasian, C.~Strickland-Constable, and Y.~Zhang, ``On symmetries and
  dynamics of exotic supermultiplets,''
  \href{http://arxiv.org/abs/2007.08888}{{\ttfamily arXiv:2007.08888
  [hep-th]}}.

\bibitem{Tanii:1984zk}
Y.~Tanii, ``{$N=8$} supergravity in six dimensions,''
  \href{http://dx.doi.org/10.1016/0370-2693(84)90337-X}{{\em Phys. Lett. B}
  {\bfseries 145} (1984) 197--200}.

\bibitem{Curtright:1980yk}
T.~Curtright, ``Generalized gauge fields,''
\href{http://dx.doi.org/10.1016/0370-2693(85)91235-3}{{\em Phys.Lett.}
  {\bfseries B165} (1985) 304}.

\bibitem{Pasti:1996vs}
P.~Pasti, D.~P. Sorokin, and M.~Tonin, ``On {L}orentz invariant actions for
  chiral $p$-forms,'' \href{http://dx.doi.org/10.1103/PhysRevD.55.6292}{{\em
  Phys. Rev.} {\bfseries D55} (1997) 6292--6298},
\href{http://arxiv.org/abs/hep-th/9611100}{{\ttfamily arXiv:hep-th/9611100}}.

\bibitem{Sen:2019qit}
A.~Sen, ``Self-dual forms: {A}ction, {H}amiltonian and compactification,''
  \href{http://dx.doi.org/10.1088/1751-8121/ab5423}{{\em J. Phys. A} {\bfseries
  53} no.~8, (2020) 084002}, \href{http://arxiv.org/abs/1903.12196}{{\ttfamily
  arXiv:1903.12196 [hep-th]}}.

\bibitem{Mkrtchyan:2019opf}
K.~Mkrtchyan, ``On covariant actions for chiral $p-$forms,''
  \href{http://dx.doi.org/10.1007/JHEP12(2019)076}{{\em JHEP} {\bfseries 12}
  (2019) 076}, \href{http://arxiv.org/abs/1908.01789}{{\ttfamily
  arXiv:1908.01789 [hep-th]}}.

\bibitem{West:2001as}
P.~C. West, ``{${{E}}_{11}$ and {M} theory},''
  \href{http://dx.doi.org/10.1088/0264-9381/18/21/305}{{\em Class. Quant.
  Grav.} {\bfseries 18} (2001) 4443--4460},
\href{http://arxiv.org/abs/hep-th/0104081}{{\ttfamily arXiv:hep-th/0104081
  [hep-th]}}.

\bibitem{Hull:2001iu}
C.~Hull, ``Duality in gravity and higher spin gauge fields,''
  \href{http://dx.doi.org/10.1088/1126-6708/2001/09/027}{{\em JHEP} {\bfseries
  0109} (2001) 027},
\href{http://arxiv.org/abs/hep-th/0107149}{{\ttfamily arXiv:hep-th/0107149
  [hep-th]}}.

\bibitem{Cremmer:1980gs}
E.~Cremmer, ``Supergravities in 5 dimensions,'' in {\em Superspace and
  supergravity : proceedings}, S.~Hawking and M.~Rocek., eds.
\newblock Cambridge Univ. Press, 1980.
\newblock Nuffield Gravity Workshop, Cambridge.

\bibitem{Bekaert:1998yp}
X.~Bekaert and M.~Henneaux, ``Comments on chiral $p$ forms,''
  \href{http://dx.doi.org/10.1023/A:1026610530708}{{\em Int. J. Theor. Phys.}
  {\bfseries 38} (1999) 1161--1172},
  \href{http://arxiv.org/abs/hep-th/9806062}{{\ttfamily arXiv:hep-th/9806062}}.

\bibitem{Henneaux:2019zod}
M.~Henneaux, V.~Lekeu, and A.~Leonard, ``A note on the double dual graviton,''
  \href{http://dx.doi.org/10.1088/1751-8121/ab56ed}{{\em J. Phys. A} {\bfseries
  53} no.~1, (2020) 014002}, \href{http://arxiv.org/abs/1909.12706}{{\ttfamily
  arXiv:1909.12706 [hep-th]}}.

\bibitem{Bonezzi:2019bek}
R.~Bonezzi and O.~Hohm, ``Duality hierarchies and differential graded {L}ie
  algebras,'' \href{http://arxiv.org/abs/1910.10399}{{\ttfamily
  arXiv:1910.10399 [hep-th]}}.

\end{thebibliography}

\providecommand{\href}[2]{#2}\begingroup\raggedright\endgroup

\end{document}